\documentclass[12pt,letterpaper]{article}

\usepackage{arxiv}

\usepackage{hyperref}

\usepackage[table]{xcolor}

\usepackage{array}

\usepackage{amsfonts,amsmath,amssymb,amsthm}
\usepackage[square, numbers]{natbib}
\usepackage{graphicx}
\usepackage[english]{babel}
\usepackage{tikz}
\usetikzlibrary{positioning}
\usepackage[T1]{fontenc}    %
\usepackage{hyperref}       %
\usepackage{url}            %
\usepackage{booktabs}       %
\usepackage{float}
\usepackage{booktabs}
\usepackage{graphicx}
\usepackage{mathtools}
\mathtoolsset{showonlyrefs}

\usepackage{booktabs}
\usepackage{multirow}
\usepackage{subcaption}
\usepackage{caption}

\usepackage{ifthen}

\graphicspath{{./PaperFigures/}}

\newtheorem{theorem}{Theorem}[section]
\newtheorem{lemma}[theorem]{Lemma}
\newtheorem{definition}{Definition}[section]
\newtheorem{assumption}{Assumption}[section]
\DeclareMathOperator{\sign}{sign}
\DeclareMathOperator{\var}{var}
\DeclareMathOperator{\cov}{cov}
\begin{document}

% Title must be 250 characters or less.

\title{Balance correlations, agentic zeros, and networks: The structure of 192 years of war and peace} % Please use "sentence case" for title and headings (capitalize only the first word in a title (or heading), the first word in a subtitle (or subheading), and any proper nouns).

%\newline
% Insert author names, affiliations and corresponding author email (do not include titles, positions, or degrees).

\author{David Dekker\\
Edinburgh Business School\\
Heriot-Watt University\\
Edinburgh, United Kingdom\\
\texttt{d.dekker@hw.ac.uk}
\And
David Krackhardt\\
Heinz College and Tepper Business School\\
Carnegie Mellon University\\
Pittsburgh, PA, U.S.A.\\
\texttt{krack@cmu.edu}
\And
\And Patrick Doreian\\
Faculty of Social Sciences\\
University of Ljubljana\\ 
Ljubljana, Slovenia\\
and\\
University of Pittsburgh\\
Pittsburgh, PA, U.S.A.\\
\texttt{pitpat@pitt.edu}\vspace{-5ex}
\And 
Pavel N. Krivitsky\\
University of New South Wales\\ 
Sydney, Australia\\\vspace{1ex}
\texttt{p.krivitsky@unsw.edu.au}
\\
\\
\\
\textit{These authors contributed equally to this work.}
}

\maketitle

\begin{abstract}
 Social network extensions of Heider's classic balance theory have led to a plethora of adaptations that have often been inconsistent with Heider and with each other. Structural balance theory as one of the branches of balance theory has mainly focused on graph partitioning, thereby assuming, for example, homogeneity in balance-driven behavior of nodes over time. We present a general model and formal notation for these social network extensions that permit social scientists to be more explicit about their balance theoretic statements, including testing of the behavioral assumptions. Specifically, we formulate statements as a comparison of two conditional probabilities of a tie,  $Ego\stackrel{q}{\text{---}}Alter$ , where the conditionals are defined by the 2-path relations $Ego\, \stackrel{r}{\text{---}}\,X\,\stackrel{s}{\text{---}}\,Alter$ and the negation, $\neg (Ego\,\stackrel{r}{\text{---}}\,X\,\stackrel{s}{\text{---}}\,Alter)$.
%  Ego\,\text{--}\,X\,\text{--}\,Alter:
% $$P(Ego\stackrel{q}{\text{---}}Alter \ | \ Ego\stackrel{r}{\text{---}}X\stackrel{s}{\text{--% -}}Alter)\quad  > \quad P(Ego\stackrel{q}{\text{---}}Alter \ | \ \neg (Ego\stackrel{r}
%  {\text{---}}X\stackrel{s}{\text{---}}Alter))$$
The key here is that \textit{q}, \textit{r} and \textit{s}, are indices that identify elements in a set of  mutually exclusive and exhaustive relations. In other words, this approach releases researchers of the confining assumption of a signed graph dichotomy. Here
we identify neutral or non-ties as distinct from negative ties by converting a signed graph to be a restricted multigraph composed of three mutually exclusive and exhaustive relations: positive ties, negative ties, and neutral-ties.  Drawing on the work on Transitivity Correlation models, we develop a set of straightforward  descriptive statistics to measure the prevalence for any stipulated balance configuration in a network by correlating the presence of a particular tie with the corresponding conditional 2-paths that surround that tie. Two major advantages are that such balance correlations can be compared directly even if network sizes and densities differ, and that specific (un)balance behaviors can be evaluated. We  apply this approach to assess network-level balance in a large data set consisting of friendly vs hostile relations between countries from 1816 to 2007. We find strong evidence  for one of the four classic Heiderian balance theory predictions, and virtually no evidence in support of the unbalanced predictions.  However, we do find stable and surprising evidence that the ``neutral'' ties are important in balancing the relations among nations. Results further suggest that prevalence of balance driven behavior varies over time, and in fact other triadic motivated behaviors prevail among countries in certain eras.
   
\end{abstract}

\section{Introduction}
Balance theory prevails as one of the most widespread theories of social behavior to emerge in the past 70 years.  With its roots in psychology \cite{Heider1958}, it is widely used in sociology \cite{moody1986balance, rawlings2017structural},
political science \cite{Krasner1985}, 
social psychology \cite{Newcomb1953}, and anthropology \cite{Hage79}.  But, perhaps no field has invested more in balance theory than social networks.  Whether these networks are studied by social scientists \cite{doreian2001pre,feld1981focused}, statisticians \cite{snijders2001statistical, fienberg1985statistical}, or physicists \cite{kirkley2019balance}, all appeal to balance theory as an explanation or motivation for actor behavior in their research.

One problem with its ubiquitous presence, however, would be that balance theory has taken on different meanings and forms as it has been adapted by the various branches of the study of human behavior, especially those studying social networks. Different network scholars have reinterpreted balance theory in such different ways that it is almost unrecognizable by those who founded the theory itself.  

Most notably we want to address the gap that has risen since \cite{CartwrightHarary1956} proved a result on network partitioning as a consequence of balance seeking behavior of nodes. This shifted research away from certain crucial elements in balance theory as it was originally conceived. Already, \cite{opp_balance_1984} noticed different strands of balance theory had come about. Where social network theory mostly focuses on structural balance theory as an explanation for group formation and partitioning in networks, the social psychology strand considers diversity in nodal affect and behavior underlying balance processes \cite{opp_balance_1984, doreian_evolution_2004, estrada_rethinking_2019}. What is not considered in neither branch is the prevalence of such behavior in the whole signed network. This is fundamental to understanding the micro-macro link, and this paper explores this issue for the network of international relations over nearly 200 years. Measuring and testing behavioral assumptions of balance theory requires to release the dichotomous relation restriction, and hence requires another more general model.

The purpose of this paper is to provide a formal approach to the study of balance in networks that is  precise yet flexible,  simple yet sophisticated, and innovative yet hones even more closely to Heider's original theoretical logic.  In a major departure from the dominant research agenda, we look beyond the traditional dichotomous perspective that the only ties of interest are the positive and negative ones.  Specifically, we ask the question, what role do agentic zeros (neutral relations) play in the consideration of balanced states in a network?   Our approach permits researchers in all fields to better specify and understand balance theory's implications and predictions under such expanded conditions.

\section{Balance in a new theoretical frame}

\subsection{Background}
Heider \cite{heider1944social, Heider1946, Heider1958} originally defined balance as a function of the relationships between three entities: a person (P), another person (O), and an inanimate object (X). Each relationship between these three entities was characterized as either positive or negative.  If the product of the signs is positive, the triad is deemed to be balanced; if the product is negative, the triad is deemed to be unbalanced. 
  This yields eight combinations of possible tie patterns. The central premise in Heider's balance theory is that P strives to attain balance due to adapting the relation to O, such that the triad becomes in any of the 'Balanced'-states shown in Table~\ref{table:HeiderBalance}.

%\centerline{+ + +, + + -, + - +, + - -, - + +, - + -, - - +, and - - -.}

\begin{table}[b]
\caption{{Heider's P\,\text{--}\,O\,\text{--}\,X designation.}}
\centering
\begin{tabular}{ c c c c} 
 \toprule
 P\,\text{--}\,O & P\,\text{--}\,X &O\,\text{--}\,X & Balance state\\
 \cmidrule(lr){1-3} \cmidrule(lr){4-4}
$+$ & $+$ & $+$ & Balanced \\ 
$+$ & $-$ & $-$ & Balanced \\ 
$-$ & $+$ & $-$ &  Balanced \\ 
$-$ & $-$ & $+$ & Balanced \\ 
$+$ & $+$ & $-$ & Unbalanced \\ 
$+$ & $-$ & $+$ &  Unbalanced \\ 
$-$ & $+$ & $+$ &  Unbalanced \\ 
$-$ & $-$ & $-$ &  Unbalanced \\ 
\bottomrule
\end{tabular}
\label{table:HeiderBalance}
\end{table}

Newcomb \cite{Newcomb1953} was among the first to  apply Heider's theory exclusively to a triad of people. By identifying ``X" as a third individual, rather than an object as Heider had done, he demonstrated how we could use balance among three people to explain many relational social dynamics.  
An important extension of this idea was developed by Cartwright and Harary \cite{CartwrightHarary1956}, who proposed formal mathematical statements about balance properties of the network \textit{as a whole}, not just isolated triples of actors. The triple was considered an unordered set without any special consideration given to one node over another --- i.e., it was an unlabeled triple. However, this has obfuscated the focus on the role of ``P'' as the actor/agent of interest in Heider's work. Indeed, the labels in ``P\,\text{--}\,O\,\text{--}\,X'' were replaced with far less discriminating labels ${i, j, k}$.  This became the standard for one branch of balance theory \cite{opp_balance_1984, doreian_evolution_2004, opp_structural_2021} (see those for description of different branches), the concept and measure of triad census as introduced by \cite{HollandLeinhardt79} %Holland and Leinhardt (Holland \& Leinhardt, 1978) 
and continues to this day \cite{dinh_enhancing_2023_2,aref_measuring_2018, aref_balance_2019, estrada_walk-based_2014, estrada_rethinking_2019}.

Rapaport \cite{Rapoport1963} provided an important theoretical contribution by translating these technical and mathematical elements  into simple statements of social relations that has inspired much of the subsequent work in this area.  Specifically, each of these eight configurations has a social interpretation that could be construed to give these balance rules an intuition in concrete, everyday terms like friendship and enemy.  These  eight theoretical statements corresponding to Heider's P\,\text{--}\,O\,\text{--}\,X model are in Table~\ref{tab:balunbal}.

\begin{table}
\caption{{Putting balance theory in words.}}
\centering
\begin{tabular}{ c c }
\toprule
Balanced & Unbalanced \\
\cmidrule(lr){1-1} \cmidrule(lr){2-2}
 A friend of a friend is a friend. & A friend of a friend is an enemy. \\ 
 A friend of a enemy is an enemy. & A friend of an enemy is a friend.  \\  
 An enemy of an enemy is a friend. & An enemy of a friend is a friend. \\
 An enemy of a friend is an enemy. & An enemy of an enemy is an enemy.
 \\
 \bottomrule
\end{tabular}
\label{tab:balunbal}
\end{table}

While such statements make the Heiderian predictions easy to understand, these also are inherently overly demanding. For example, no one really believes that \textit{everytime} someone encounters a friend of a friend that she will establish a friendship with that person.  Rather, a more reasonable prediction from this first Heiderian principle is that a friend of a friend is \textit{more likely} to be a friend than if that person is not a friend of a friend.  That is, the 2-path relation, friend-of-a-friend, creates a favorable \textit{context} in which a friendship is more likely to form \textit{than without that context}. Unfortunately, measures of the Degree of Balance, usually don't allow to distinguish between these different types of (un)balance behaviors \cite{CartwrightHarary1956, harary_measurement_1959, dinh_enhancing_2023, aref_measuring_2018, estrada_walk-based_2014}. 

\subsection{Balance theory as a set of probability statements}
We will throughout this paper replace Heider's P\,\text{--}\,O\,\text{--}\,X terms with corresponding network terms.  We start by noting that the only relationship that Heider specifically designated as  interpersonal was the relationship between P and O. And the question posed by the model is what relation P would choose to have with O, given the association both have with X. This X provides a contextual element that influences P's choice.  Consistent with the social network literature, we refer to this P$\,\text{--}\,$O relationship as an Ego $\,\text{--}\,$Alter relationship, wherein Ego ($E$) is choosing a relationship with Alter ($A$).  We retain the label ``X'' to refer to a contextual third party.  With this minor amendment, we can formalize the first Heiderian ``a friend of a friend is a friend'' principle with the following conditional inequality:
\begin{equation} \label{prob_fff_bal_a}
     P(E \stackrel{f}{\text{---}}A \ | \ E\stackrel{f}{\text{---}}X\stackrel{f}{\text{---}}A)  > P(E\stackrel{f}{\text{---}}A \ | \ \neg  (E\stackrel{f}{\text{---}}X\stackrel{f}{\text{---}}A))
    % \hss   
\end{equation}
or, equivalently,
 \begin{equation} \label{prob_fff_bal_b}
P(E \stackrel{f}{\text{---}}A \ | \ E\stackrel{f}{\text{---}}X\stackrel{f}{\text{---}}A)  - P(E\stackrel{f}{\text{---}}A \ | \ \neg  (E\stackrel{f}{\text{---}}X\stackrel{f}{\text{---}}A)) > 0
\end{equation}
where $\stackrel{f}{\text{---}}$ indicates a friendship tie.  Similar conditional inequalities can be made of each of the other three primary Heiderian predictions about balanced relations.  In cases of \textit{unbalanced } relations, such as ``a friend of an enemy is a friend,'' then the prediction is that the 2-path relation,  being a friend of an enemy, should lead to relatively infrequent friendships.  Such unbalanced statements would be similarly described as in \ref{prob_fff_bal_a}, except that the inequality is reversed:
\begin{equation} \label{prob_fff_bal_c}
 P(E \stackrel{f}{\text{---}}A \ | \ E\stackrel{e}{\text{---}}X\stackrel{f}{\text{---}}A)  < P(E\stackrel{f}{\text{---}}A \ | \ \neg  (E\stackrel{e}{\text{---}}X\stackrel{f}{\text{---}}A))
\end{equation}
or, equivalently,
 \begin{equation} \label{prob_fff_bal_d}
      P(E \stackrel{f}{\text{---}}A \ | \ E\stackrel{e}{\text{---}}X\stackrel{f}{\text{---}}A)  - P(E\stackrel{f}{\text{---}}A \ | \ \neg  (E\stackrel{e}{\text{---}}X\stackrel{f}{\text{---}}A)) <0 \
\end{equation}
where $\stackrel{f}{\text{---}}$ indicates a symmetric friendship tie and  $\stackrel{e}{\text{---}}$ indicates a symmetric enemy relation.

At the network level, we can frame this prediction as a statement about the relative likelihood that the predicted tie would appear given that it is facing a contextual 2-path relation.  That is, in the network as a whole, we can cross-tabulate the number of occurrences of a tie type with the number of occurrences of the specified 2-path relation, shown in Table~\ref{table:xtab_tie_vs_2step}.

\begin{table}
\caption{\label{table:xtab_tie_vs_2step}{Cross-tabulation of Ego\,\text{--}\,Alter ties with 2-path triples, Ego\,\text{--}\,X\,\text{--}\,Alter.}}
\centering
\begin{tabular}{ r c  c  c }
\toprule
 &   \multicolumn{3}{c}{\quad Is there a 2-path Ego\,\text{--}\,X\,\text{--}\,Alter?} \\
 %\hline
   &     & 1=Yes &  0=No\\
   \midrule
  \multirow{3}{3cm}{Is there a tie Ego\,\text{--}\,Alter?}&   1=Yes  &  $a$ & $b$ \\
 % \hline
 &    &  & \\
   &   0=No  &  $c$ & $d$ \\
\bottomrule
\end{tabular}
% \vskip.1in
\begin{flushleft}
    \textbf{Note:} Co-occurrence type counts are given by $a$, $b$, $c$, and $d$.
\end{flushleft}  
\end{table}

 If we assume a uniform and random draw from all $\{i,j,k\}$ triples in the network, then we can directly calculate the relevant conditional probabilities from this 2x2 table.
The first conditional in Eq~\ref{prob_fff_bal_a} and Eq~\ref{prob_fff_bal_b},
the probability that a friendship between Ego and Alter exists, given that there is an X who is also a friend of Ego and a friend of Alter, is calculated as ${a}/{(a+c)}$.  The second conditional in each equation is calculated as ${b}/{(b+d)}$.  Thus, the difference between these two fractions 
$({a}/{(a+c)}) - ({b}/{(b+d)})$
represents the extent to which the particular balance theory prediction is true.  The larger the difference, the stronger the association given the conditional 2-path.  If the relation is unbalanced, such as described in Eq. 3 and 4, then the difference between these two fractions would be negative.  The larger the negative value, the stronger the effect of the 2-path in suppressing the existence of the Ego\,\text{--}\,Alter tie.  

Conversely, if $({a}/{(a+c)}) - ({b}/{(b+d)})$ is positive, then it will also be the case that the difference between the corresponding row percentages will be positive 
(i.e., $({a}/{(a+b)}) - ({c}/{(c+d)}) >0$).  This represents the symmetric claim  that if the Ego\,\text{--}\,Alter tie exists, then the probability of the designated 2-path Ego\,\text{--}\,X\,\text{--}\,Alter tie will be greater than if the Ego\,\text{--}\,Alter tie does not exist.  While these differences will not in general exactly equal each other, they will always be of the same sign.  Moreover, the size of these differences will correspond to the strength of the particular Heiderian balance principle.  

Putting all this together, we can describe the size of the balance theory effect by taking the geometric mean of these two differences in probabilities:
$$\sqrt{\left( \frac{a}{a+c} - \frac{b}{b+d}\right)\left( \frac{a}{a+b} - \frac{c}{c+d}\right)}. $$

Finally, we draw on earlier work \cite{dekkeretal2019} to note that this geometric mean is equal to the Pearson correlation between  the vector of their corresponding Ego\,\text{--}\,X\,\text{--}\,Alter 2-path ties for all $\{i,j,k\}$ triples and the corresponding vector of their associated Ego\,\text{--}\,Alter ties.  That is,
\begin{equation} \label{bal_cor_eqn}
\rho=s\sqrt{\left( \frac{a}{a+c} - \frac{b}{b+d}\right)\left( \frac{a}{a+b} - \frac{c}{c+d}\right)} ,
\end{equation}
where
\begin{equation} \label{eq:sgn}
    s=\sign\left( \frac{a}{a+c} - \frac{b}{b+d}\right). 
\end{equation}

The sign function in Eq~\ref{eq:sgn}  differentiates between balanced and unbalanced predictions.  If ${a}/{(a+c)} > {b}/{(b+d)}$, then the correlation between the presence of the predicted Ego--Alter tie and its conditional 2-path will be positive; thus, the positive root of this square root function is used, indicating that the Ego\,\text{--}\,Alter tie probability is enhanced in the presence of the Ego\,\text{--}\,X\,\text{--}\,Alter 2-path.  If ${a}/{(a+c)} < {b}/{(b+d)}$, then this correlation will be negative; thus, the negative root of this square root function is used, indicating that the Ego\,\text{--}\,Alter tie probability is suppressed in the presence of the Ego\,\text{--}\,X\,\text{--}\,Alter 2-path.  

In other words, the Pearson product-moment correlation between observed Ego\,\text{--}\,Alter ties and their associated Ego\,\text{--}\,X\,\text{--}\,Alter 2-path ties directly measures the observed prevalence of balance (or unbalance) in the network as a whole.

\subsection{The critical role of (agentic) non-ties}
Another restriction in Heider's model  is that it assumes all relations are either positive or negative,  not neutral. On the other hand, in reality many people are going to have neutral sentiments (neither positive nor negative) about each other. While such non-ties are often seen as part of any natural network, they are often treated as  ``missing data''  and are excluded from any predictions 
\cite{CartwrightHarary1956}.  Or,  by some conventions,  as noted by \cite{rawlings2017structural},  ``... the absence of a positive sentiment implies the presence of a negative sentiment,''  and as such a lack of a tie may be deemed to  represent a negative tie. 

In a major departure from this tradition, we explicitly consider neutral relations as agentic or social choices, just as much as positive and negative relations.  There are many reasons that neutral or non-relationship ties should be considered a choice within a balance theory context.   For example, in a highly polarized and conflict-ridden environment, Ego may have positive sentiments toward X, but  Ego also observes that Alter and X have an intense dislike for each other.  If Ego were to develop a positive relation with Alter, even if it were to serve Ego's interests to do so, this would create an unbalanced triple.  One way of escaping this uncomfortable position is to establish a neutral or non-relationship with Alter, rather than taking sides in the dispute.  That is, Ego may take the ``Switzerland'' option of staying neutral in the spat, in order to preserve her options to work with Alter in the future. 

Strategic neutrality is not the only reason for considering neutral relations in balance theory. Within the blockmodeling perspective it has been noted frequently how zero blocks (groups of people who are discouraged from relating to each other or to other groups) can be found as a way to preserve social order \cite{whiteboormanbreiger1976,pachuckibreiger2010}. The cost of forming ties under such conditions could be substantial in a social situation.  Understanding how these non-ties play a role in balancing triadic relations is explicitly part of our model.

Yet another reason for observing null ties, even in cases where relationships to all other actors are evaluated is a limitation in resources. Such restrictions prevent engagement in positive or negative ties. For example, in international relations smaller countries may not have sufficient budget or human capital to invest in embassies in each other country. Also, an isolationist approach to foreign affairs may prevent joining collective security agreements. Or negatively, countries cannot afford to go at war in case of resource dependencies. Although different ideologies, internal policies and choices of countries may lead to neutrality, the behavioral consequence is uniform, and distinct from positive and negative ties.

Thus, Ego may develop a neutral attitude towards Alter. Moreover, Ego's observation that X may have a neutral relation with either Ego or Alter (or both) could affect Ego's calculations in how to consider her tie with Alter.  We therefore maintain that any prediction stemming from balance theory about a real network must recognize this inherent character of the network and include positive, negative \textit{and neutral} ties as part of the balance processes.

\subsection{Triad signature}
Our claim, then, is that all signed graphs can be decomposed into three (sentiment) subnetworks: Positive ties, Negative ties, and Zero ties (neutral or non-ties).  These three types of relations define a restricted multigraph on any network  $\mathcal{G}$:
\begin{description}
\item[$\{\textbf{P}\}:=$] the set of ($i,j$) unordered pairs designating all ties where $i$ has a \textit{positive} tie with $j$;

\item[$\{\textbf{N}\}:=$] the set of ($i,j$) unordered pairs designating all ties where $i$ has a \textit{negative} tie with $j$; and

\item[$\{\textbf{Z}\}:=$] the set of ($i,j$) unordered pairs designating all ties where $i$ does \textit{not} have a positive tie nor a negative tie with $j$, which we indicate as a \textit{neutral} or ``zero'' tie.

\end{description}

We ignore self-ties $(i,i)$ and  hence these will not be in these three sets. By convention, we represent these three sets in square matrix form as follows:
\[
\textbf{P}_{ij} = 
\begin{cases}
    1 & \text{if  ($i,j$)} \ \in \{\textbf{P}\}\\
    0              & \text{otherwise},
\end{cases}
\quad
\textbf{N}_{ij} = 
\begin{cases}
    1 & \text{if  ($i,j$)} \ \in \{\textbf{N}\}\\
    0              & \text{otherwise},
\end{cases}
\quad
\textbf{Z}_{ij} = 
\begin{cases}
    1 & \text{if  ($i,j$)} \ \in \{\textbf{Z}\}\\
    0              & \text{otherwise.}
\end{cases}
\]
These sets are, by definition, mutually exclusive and exhaustive in the network; thus, their sum, as well as their union, forms a complete and simple graph on the full set of nodes in $\mathcal{G}$.  These three networks (\textbf{P, N} and \textbf{Z}) henceforth will be referred to as the set of \textit{primitive relations} in $\mathcal{G}$. 

Given the three possible relations in any ($i,j$) pair, then there are $3\times3=9$ possible configurations in any 2-path $i\text{-} k\text{-} j$.  For each 2-path configuration, there are  three relations between $i$ and $j$ that complete the triple, yielding $3^3 = 27$ configurations possible. Each of these may be designated with a 3-tuple identifier.  We will use the following rules to identify each unique possible configuration:

Given any ordered triple of actors in graph $\mathcal{G}$, we designate three roles, one ascribed to each of the three actors: Ego, the focal person to whom the balance consideration is paramount; Alter, the person whom Ego is potentially establishing a relationship with; and Context, the person with whom both Ego and Alter have a relation that defines the context that determines whether the $i\text{-} j$ relation is in balance or not.  We will heretofore abbreviate these role references as ``E'' for Ego, ``A'' for Alter, and we retain Heider's ``X'' for the Contextual third party.  Any labeled triple, then, has three tie values, designated by the three pairs of the three nodes. Each  pair will be accorded a value corresponding to one of the three types of ties (Positive, Negative, or Zero  --- from hereon out  referred to as $p$ ties, $n$ ties and $z$ ties, respectively).  Using this notation, any and all configurations of ties amongst any three nodes may be designated by assigning the letter $p$, $n$ or $z$  to each of the three  pairs.  The first letter will designate the relation between Ego and Alter; the second will designate the relation between Ego and X; and the third letter will designate the relation between Alter and X (see Table~\ref{triad.signature.position}).  

\begin{table}
\caption{{Triad signature position designation}}
\centering
\begin{tabular}{c c l } 
\toprule
 Position & &Tie Designation\\
    \cmidrule{1-1}\cmidrule{3-3}
 1 & & $E\stackrel{q}{\text{---}} A$  (Ego with Alter) \\

2 & & $E\stackrel{r}{\text{---}} X$ (Ego with X) \\

3 & & $A\stackrel{s}{\text{---}} X$ (Alter with X) \\
\bottomrule
\end{tabular}
\begin{flushleft}
 \textbf{Note:} The \textit{triad signature} is a 3-tuple. Each edge, $q, r,$ and $s$, designates an undirected tie of type $p, n,$ or $z$ among the triple labeled Ego, Alter and X.   
\end{flushleft}
\label{triad.signature.position}
\end{table}

For reasons that will become clear later, we will separate the first letter in this 3-tuple from the remaining two letters with a period.  Thus, for example, the 3-tuple \textit{<p.nz>} would indicate that Ego has a positive tie with Alter,  Ego has a negative tie with X, and Alter has a zero tie (non-tie) with X.  Heider's ``the enemy of an enemy is a friend'' prediction would be represented as  \textit{<p.nn>}.  These three letters, in the designated order, provide a \textit{triad signature},  which permits the unique designation of each of the $3^3\  =27$ possible configurations that Ego might face.

While any ordering of these letters is possible, we have chosen this arbitrary ordering because it allows one to see some of the principles behind Heider's theoretical predictions.  In fact, the consistent application of this order of these three elemental relations in the signature provide three important features: 1) The order identifies a specific structural configuration, one of the possible 27; 2) The order represents specific and formal \textit{predictive} statements from balance theory; and 3) The order naturally leads to a precise calculation of the balance correlation measure of the network-level of balance specifically  designated by the signature. We will discuss each of these features in turn below.

\subsubsection{Identifying the 27 triad signatures}  
\begin{figure}
\caption{The 27 Triad Signatures.}
% \vspace{-0.05cm}
\centering

\foreach \svalue in {p,n,z} {
  \foreach \qvalue in {p,n,z} {
    \foreach \rvalue in {p,n,z} {
      \begin{subfigure}{.3\textwidth}
        \centering
        \begin{tikzpicture}[scale=.9]
           %\vspace{1cm}
          % Define node styles with smaller size and font size
          \tikzstyle{node} = [circle, draw, fill=gray!10, text centered, minimum size=0.4cm, font=, font=\footnotesize]
          \tikzstyle{edge} = [thick]
      
          % Nodes
          \node (ego) at (0, 0) [node] {E};
          \node (alter) at (2*0.55, 0) [node] {A};
          \node (x) at (0.55, 2*0.55) [node] {X};

          % Determine edge color and pattern based on label values
          \ifthenelse{\equal{\rvalue}{p}}{
            \draw[edge, black] (ego) -- (alter) node[midway, below, black, font=\footnotesize] {p};
          }{
            \ifthenelse{\equal{\rvalue}{n}}{
              \draw[edge, gray!50, dashed] (ego) -- (alter) node[midway, below, black, font=\footnotesize] {n};
            }{
              \ifthenelse{\equal{\rvalue}{z}}{
                \draw[edge, white] (ego) -- (alter) node[midway, below, black, font=\footnotesize] {z};
              }{}
            }
          }

          \ifthenelse{\equal{\svalue}{p}}{
            \draw[edge, black] (ego) -- (x) node[midway, left, black, font=\footnotesize] {p};
          }{
            \ifthenelse{\equal{\svalue}{n}}{
              \draw[edge, gray!50, dashed] (ego) -- (x) node[midway, left, black, font=\footnotesize] {n};
            }{
              \ifthenelse{\equal{\svalue}{z}}{
                \draw[edge, white] (ego) -- (x) node[midway, left, black, font=\footnotesize] {z};
              }{}
            }
          }

          \ifthenelse{\equal{\qvalue}{p}}{
            \draw[edge, black] (alter) -- (x) node[midway, right, black, font=\footnotesize] {p};
          }{
            \ifthenelse{\equal{\qvalue}{n}}{
              \draw[edge, gray!50, dashed] (alter) -- (x) node[midway, right, black, font=\footnotesize] {n};
            }{
              \ifthenelse{\equal{\qvalue}{z}}{
                \draw[edge, white] (alter) -- (x) node[midway, right, black, font=\footnotesize] {z};
              }{}
            }
          }
          % Add the label "<r.qs>" under each triad
          \node[left=1.5cm, font=\footnotesize] at (x) {<$\rvalue.\svalue\qvalue$>};
        \end{tikzpicture}
      \end{subfigure}
      \vspace{0.2cm}
    }
  }
}
\vspace{-0.95cm}
    \begin{flushleft}
        \textbf{Note:} Each signature is given by any combination of \textit{p}, \textit{n} and \textit{z}, in angle brackets. A solid line indicates a positive tie; a dashed line indicates a negative tie; no line indicates a zero-tie. Node labels indicate roles of Ego (E), Alter (A), and Contextual (X).
    \end{flushleft}
    \label{fig:signatures}
\end{figure}
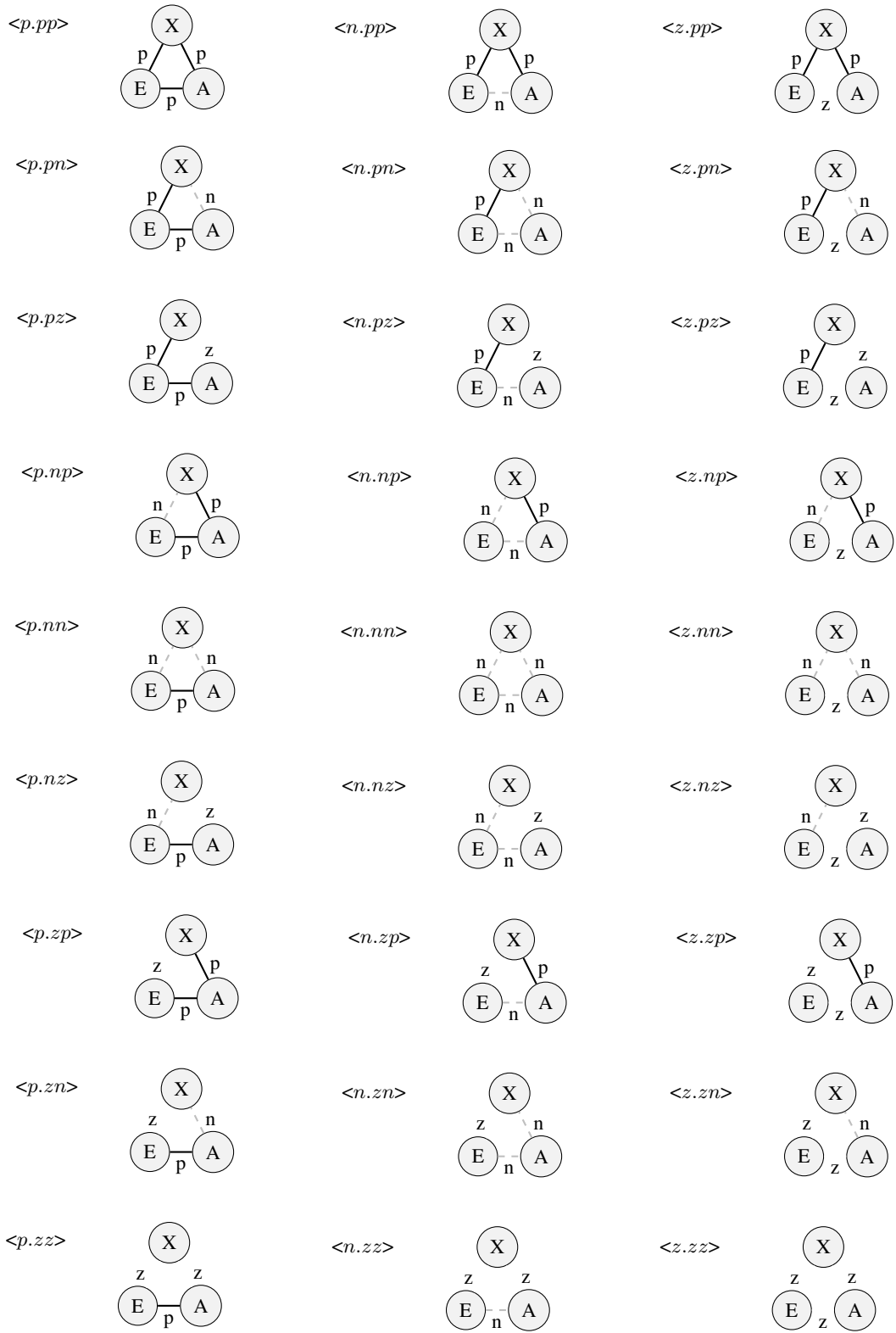

The 27 possible labeled configurations are provided in  Fig~\ref{fig:signatures}.  Positive ties (\textit{p}) are represented in black; negative ties (\textit{n}) are represented in dashed red lines; zero-ties (\textit{z}) or neutral ties are blank, and are indicated with a letter ``z'' to emphasize they are treated as ties. 
Node labels refer to the Ego (E), Alter (A), and Context (X) roles in each triad.  As we will discuss later, at the network level some of these triads will be mathematically redundant. Nonetheless, in order to keep the theoretical stories clear, it is still important to retain all 27 triad signatures. 

As an interesting extension of this last point, we note that many of these labeled configurations are isomorphic, yet their theoretical claims are still distinct. For example, \textit{<p.pz>} is isomorphic with \textit{<z.pp>}, but each is a different claim on the experience that Ego faces.  In the former case, Ego is the central character, the bridge between two friends who are neutral toward each other. In the \textit{<z.pp>} case, however, Ego is neutral to Alter, even though his friend (X) likes Alter.  How each of these two configurations may play out in Ego's experience from a balance theory perspective could be quite different, depending on the options Ego experiences.  In the \textit{<z.pp>} case, Ego could actively close the triple with a positive tie to Alter, creating a \textit{<p.pp>} triple, which would be attractively balanced in Heider's view.  But, in the \textit{<p.pz>} case, Ego has less agency, since the neutral tie is between Alter and X.  Thus, these two isomorphic structures represent entirely different balancing experiences and options from Ego's viewpoint.

%\subcaptionsetup{labelformat=empty}

\subsubsection{Triad signature as a predictive statement} Each triadic signature also operates as a unique theoretical predictive statement regarding balance processes.  To see how this is so, we return to the  predictive statement that Heiderian balance can make: If balance theory holds, then Ego should be motivated to create balance in its ties and to alter relations that create unbalanced situations.  In a triadic context, the probability that Ego will actively choose to engage in a tie (either \textit{p}, \textit{n} or \textit{z}) with Alter will depend on the 2-path relation through X.  In the present model, the first letter in the triad signature represents this choice faced by Ego.    This separate role is visually enhanced by the placement of a period between it and the latter two letters in the signature.

To illustrate this feature, we begin with a list of all eight classic Heiderian balance-theory statements, along  with our triadic signature notation in Table \ref{table:rapaport and signatures}. 
The core prediction in Heider's model is that if a triple is unbalanced, then this creates psychological tension and discomfort.  Thus, an unbalanced triple should not occur, or if it does occur, an attempt will be made by Ego to balance it by reconsidering her relations with Alter to make it balanced.  On the other hand, if a triple is balanced, it will induce a sense of equanimity on Ego's part and is more likely to be stable and remain in place. 
Viewed in this way,  each \textit{balanced} signature easily translates to a corresponding prediction from balance theory. Consider the following decomposition of the triadic elements.

\begin{table}
\caption{{Rapaport descriptions and the corresponding triad signature representations.}}
\centering

\begin{tabular}{ l c c l c}
\toprule
Balanced & signature & &Unbalanced & signature\\
\cmidrule{1-2} \cmidrule{4-5}
 A friend of a friend is a friend.&\textit{<p.pp>}& & A friend of a friend is an enemy.&\textit{<n.pp>}\\
 A friend of a enemy is an enemy.&\textit{<n.np>}& &A friend of an enemy is a friend.&\textit{<p.np>}\\  
 An enemy of an enemy is a friend.&\textit{<p.nn>}& &An enemy of a friend is a friend.&\textit{<p.pn>}\\
 An enemy of a friend is an enemy.&\textit{<n.pn>}& &An enemy of an enemy is an enemy.&\textit{<n.nn>}\\
\bottomrule
\end{tabular}
\label{table:rapaport and signatures}
\end{table}

Given any set of relational ties ($q$, $r$, and $s$) among Ego, Alter and X, such that Ego has a $q$-tie with Alter, Ego has an $r$-tie with X, and Alter has an $s$-tie with X, the signature of this triad \textit{<q.rs>} can be decomposed into two parts: the $q$-tie, and the 2-path $rs$-tie.  Now all balance theory statements can be framed by treating the $q$-tie  as a dependent variable, predicted by the contextual 2-path, the $rs$ component (i.e.,  $P(q)=f(rs)$).
% \end{adjustwidth}

As a general principle, if the triple \textit{<q.rs>} is \textit{balanced}, then the prediction is
\begin{equation} \label{prob.qrs.bal}
P(E\stackrel{q}{\text{---}} A \ | \ E\stackrel{r}{\text{---}} X \wedge A\stackrel{s}{\text{---}} X)\quad  > \quad P(E\stackrel{q}{\text{---}} A \ | \ \neg  (E\stackrel{r}{\text{---}} X \wedge A\stackrel{s}{\text{---}} X))
\hss.
\end{equation}
To illustrate, the Heiderian balance prediction that ``the enemy of an enemy is a friend'' can be restated more formally as ``the probability that a specific Alter will be a friend is enhanced if I have an enemy who is also the enemy of this Alter.''  This statement immediately corresponds to and  is represented by the configuration \textit{<p.nn>}, where \textit{p} is the dependent variable and $nn$ is the 2-path conditional.  The prediction is that $p$ is more likely to occur given the presence of the 2-path $nn$ than if $nn$ is not present:
 \begin{equation} \label{prob.qrs.unbal}
% \hskip.5in
P(E\stackrel{q}{\text{---}} A \ | \ E\stackrel{r}{\text{---}} X \wedge A\stackrel{s}{\text{---}} X)\quad  < \quad P(E\stackrel{q}{\text{---}} A \ | \ \neg  (E\stackrel{r}{\text{---}} X \wedge A\stackrel{s}{\text{---}} X))
\hss.
\end{equation}

Take as an example the second unbalanced statement in Table~\ref{tab:balunbal}, ``A friend of an enemy is a friend,'' represented by the configuration \textit{<p.np>}.  Since it is theoretically \textit{ unbalanced},  it is predicted to create an uncomfortable situation for Ego.  Given that the 2-path relation ``a friend of an enemy'' exists between Ego and Alter, then it is predicted that Ego will \textit{not},  or at least be less likely to, establish a positive friendship tie to Alter. The specific prediction, then, is that a \textit{p}-tie is less likely to occur given the presence of the 2-path \textit{np} than if \textit{np} is not present.

All balance theory  statements can be framed this way.  Evidence for or against these predictions, stemming directly from triad signatures, can be brought to bear on each statement to see which specific predictions hold up under scrutiny.   While this sounds straightforward, we will soon see that the issue of balance is considerably more complicated than this simple interpretation.

\subsubsection{Triad signature-induced measure of network balance: The balance correlation}

\begin{figure}
\caption{{Example of a  network illustrating the embeddedness principle}} 
\centering
\begin{tikzpicture}
  % Define node styles
  \tikzstyle{node} = [circle, draw, fill=gray!10, text centered, minimum size=0.6cm, font=\small]
  \tikzstyle{edge} = [thick]

  % Nodes
  \node (ego) at (0, 0) [node] {Ego};
  \node (node1) at (2.5, 0) [node] {1};
  \node (node2) at (1.5, 1.5) [node] {2};
  \node (node3) at (2, -2) [node] {3};
  \node (node4) at (3.5, -1.75) [node] {4};
  \node (node5) at (5, -1.5) [node] {5};
  \node (node6) at (-2, 1.5) [node] {6};
  \node (node7) at (-1, 2.75) [node] {7};
  \node (node8) at (0, 4) [node] {8};

  % Positive ties from Ego to all other nodes
  \foreach \target in {1,2,3,4,5,6,7,8}
    \draw[edge, black] (ego) -- (node\target);

  % Negative ties from node 2 to nodes 1, 6, 7, and 8
  \foreach \target in {1,6,7,8}
    \draw[edge, gray!50, dashed] (node2) -- (node\target);

  % Positive ties from node 1 to nodes 3, 4, and 5
  \foreach \target in {3,4,5}
    \draw[edge, black] (node1) -- (node\target);
\end{tikzpicture}
\begin{flushleft}
  \textbf{Note:} The black edges represent positive ties; the dashed edges represent negative ties.   The focal labeled triple of interest here is \{Ego,1,2\}, which is unbalanced (<\textit{p.pn}>). Since the Ego-2 tie is embedded in more unbalanced triples than the Ego-1 tie is,  Ego is more likely to change her relationship with node 2 as an Alter than node 1 as an Alter in trying to resolve this unbalance.  
\end{flushleft}
\label{fig:example.network.1}

\end{figure}
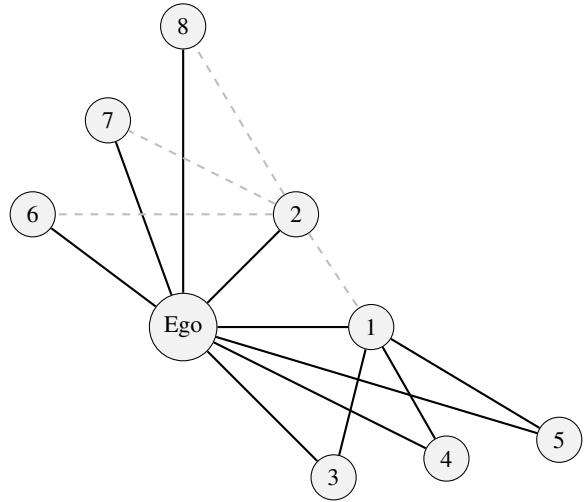

{\bf The Embeddedness Principle.}  Before we introduce this third, and most critical, function of the Triad Signature, we note one final difference between Heider's original theoretical formulation and applications to social network analysis.   
Heider himself focused on isolated and ideational triples of entities %. Each triple labeled a Person, an Other, and some contextual construct X, 
and made predictions based solely on these three entities.  In a network context, however, it is important to recognize that no labeled Ego\,\text{--}\,X\,\text{--}\,Alter combination occurs in isolation but rather is embedded in a complex structure of triples. In fact, changes in one tie can imply balance changes elsewhere in the network. Consider the case depicted in Fig~ \ref{fig:example.network.1}.  The three nodes in the middle of the graph (Ego, 1, and 2), constitute an unbalanced triple: Ego has a positive tie with Node 1, and Ego also has a positive tie with Node 2;  but, Nodes 1 and 2 have a negative tie.  Consider the specific ordered triple  where  node 1 is Alter and node 2 is the contextual X.  Ego is faced with a dilemma.  In considering whether to change her relationship with Alter (node 1), Ego finds herself embedded in several critical triples with node 1 as an Alter ((Ego,1,3), (Ego,1,4), (Ego,1,5)).  Each of these triples has the signature of \textit{<p.pp>}, and therefore these are balanced.  If Ego were to balance the (Ego,1,2) triple by switching her relationship with node 1 from \textit{<p>} to \textit{<n>}, then that would distort the balance that currently exists in these three other \textit{<p.pp>} triples.  In Heiderian terms, the balance cost would far outweigh the balance benefits for switching.

On the other hand, consider the alternate ordered triple where  Ego is now taking stock of her relationship with 2 as the Alter, in light of node 1's new role as Context (X).  With node 2 as Alter, Ego also has to consider that there are, again, several contexts ((Ego,2,6), (Ego,2,7), and (Ego,2,8)), to take into account, not just the triple with node 1 (Ego,2,1).  In this case, however, all the pertinent triples are currently unbalanced with a \textit{<p.pn>} structure.    If Ego  were to switch her relation to Alter (node 2) from \textit{<p>} to \textit{<n>}, then all four of these ordered triples would suddenly become a balanced triple \textit{<n.pn>}.  Therefore, in this case, it is easy to see that the balance forces greatly favor Ego's choice of changing her tie to Node 2 from positive to negative to fix the unbalance in her embedded network.

Not all complex contexts are as easily resolved as in this artificial example.  What needs to be underscored here, though, is that in a network context, the Ego\,\text{--}\,Alter pair does not have one sole X to consider in any balance calculation. And, the larger the number of contextual 2-paths  that constitute pre-conditions for balance, the greater the nudging force on the Ego\,\text{--}\,Alter pair to make a relationship that obeys the rules of balance.  Indeed, each Ego\,\text{--}\,Alter pair faces (n-2) triples of some configuration; some of these may be balanced, some unbalanced, and perhaps some undecided.  The Embeddedness Principle simply states that all of these triples must be considered in assessing what the balance forces are on any Ego\,\text{--}\,Alter pair.  

Taking this Embeddedness Principle into account, we now make Heider's prediction more concrete.   Since social networks provide many observations, even around a specific dyad, this affords us the opportunity to turn the prediction into a \textit{measurable probabilistic statement}, rather than simply ``the network is balanced'' or ``the network is unbalanced'', as \cite{CartwrightHarary1956} originally suggested.  We propose that balance statements are better framed with words like ``tends to'' or ``is more likely to''. 
Consider the ``friend of a friend is a friend'' prediction. 
Rather than stating ``In every case, a friend of a friend is a friend'', we argue that a better, empirically testable extension of this claim would be ``A friend of a friend will tend to be a friend.'' \textit{And the more friend-of-friend contexts one is embedded in, the more likely Ego will choose Alter as a friend}, assuming balance theory is correct.   This insight will help us explore Heider's model more precisely.   

One of the basic predictions stemming from balance theory is that, if individuals are prone to attaining and maintaining balance, then at any given time we should find that balanced relations are more prevalent in a network. That is, if balance theory actually works to influence overall structures, then we should find that the 2-paths that exist in a network should be associated with a higher %the improved 
probability of the balance-confirming direct tie from Ego to Alter.  As we observed in Eq~\ref{prob_fff_bal_a}, the more a balance-based 2-path conditional exists around any particular Ego\,\text{--}\,Alter pair, the higher the probability we would observe the predicted direct tie from Ego to Alter.

A key advantage of formalizing Heiderian predictions using triad signatures is that they make it easy to extend this correlational measure to include this embeddedness principle.  \cite{dekkeretal2019} created a correlation coefficient, the Transitivity Correlation, that assesses the extent to which the 2-path pre-transitive condition in a digraph (${i}\text{-}{k} \text{-} {j}$) is correlated with the presence of a direct tie ${i}\text{-}{j}$.  Their solution to this problem was simply to correlate the off-diagonal elements in the matrix form of $\mathcal{G}$ with the corresponding elements in matrix $\mathcal{G}^2$.  Since $\mathcal{G}^2$ contains the \textit{number} of directed 2-path ties of the form 
$i \text{-} k \text{-} j$ 
for every 
$\langle i,j \rangle $
dyad, this point biserial correlation indicates the extent to which there is an overall tendency in the digraph $\mathcal{G}$ to close a directed 2-path with a directed tie. If the correlation is strong and positive, that would indicate that the tendency towards transitive closure is increased as a function of the number of directed 2-path paths surrounding any dyad. A negative correlation would indicate that the more directed 2-path paths that surround a dyad, the \textit{less likely} it is that $i$ will be directly tied to $j$, indicating that $\mathcal{G}$ has an intransitivity propensity (such as one would find in a star-shaped graph).  A zero correlation would indicate that, controlling for the density in $\mathcal{G}$ and the number of directed 2-paths in $\mathcal{G}$, there is no overall propensity towards transitivity in $\mathcal{G}$.   

We can use this same approach to look at balance in a graph. While the solution is not as simple as correlating $\mathcal{G}$ with $\mathcal{G}^2$, we can still create a matrix of the number of contextual 2-paths that would be specified by the balance theory prediction; and we can just as easily create another matrix of ties that would be predicted by this same balance theory statement.  If these matrices are positively correlated, then that would constitute evidence that the specific balance predictions tend to hold in $\mathcal{G}$.  

This approach requires calculating for any given signed graph, $\mathcal{M}$, two matrices, 
$\textbf{M}_1$ and $\textbf{M}_2$, 
where $\textbf{M}_1$  contains the predicted Ego\,\text{--}\,Alter relation, and  $\textbf{M}_2$ contains the number of conditional 2-paths around each dyad that are being tested.  These two matrices are straightforward derivatives of the signature notation we have used thus far.  In the general case, where the triad signature describing the predicted balance relation is given by the 3-tuple \textit{<q.rs>}, 
then 
$${\textbf{M}_1} = \textbf{ Q},\ {\textbf{M}_2} =  \textbf{RS},$$
where $\bf RS $ is the inner product of the relation $\bf R$ and the  relation $\bf S$.   

A balance correlation, then, is simply the product-moment correlation between   off-diagonal elements in  $ \textbf{M}_1$ with the corresponding elements in $ \textbf{M}_2$.  Moreover, as just demonstrated, these correlations are easily generated from the signatures themselves.  For example, the \textit{<n.pn>} signature generates a balance statement that negative n-ties are induced from an abundance of 2-path pn-ties; and the balance correlation is  $\rho(\bf N, PN)$, where $\textbf{N}$ is the matrix of n-ties, and $\textbf{PN}$ is the compound word \cite{whiteboormanbreiger1976} created by the matrix product of the \textbf{P} matrix and the \textbf{N} matrix of ties .  

Balance theory predicts a positive correlation on balanced signatures and a negative correlation on unbalanced signatures.  Whether a network is balanced overall, in this view, depends on which of the possible balanced or unbalanced signatures you are looking at.  For example, within the same network, there could be a strong positive balance correlation for the predictive \textit{<p.nn>} statement but a negative correlation for the predictive \textit{<n.pn>}.  Or, perhaps we observe a positive correlation for \textit{<p.pp>} but also a positive correlation for \textit{<p.pn>} (which, according to balance theory, should be negative).  That is, balance theory may hold for some configurations but not others. 

\textbf{Phenotypical Twins.} 
As we have already noted above, there are $3^3=27$ triad signatures.  And, each of these  signatures doubles as a possible prediction and generates a correlation that permits us to evaluate this prediction.  It should be recognized, while these predictive statements and measured correlations can be generated for any of the 27 signatures, many of these signatures do not correspond to anything that Heider would claim actually stems from balance theory.  Yet, it is still possible that these empirical correlations might yield insights into structural patterns that could lead to new theoretical extensions, akin to what balance theory has produced over the past half century \cite{HollandLeinhardt79, davis1979}.  A thorough exploration of such patterns in any dataset would result in a 9 $\times$ 3 correlation matrix, where each of the three dependent relations (\textbf{Q, R,} and \textbf{S})  is potentially predicted by each of the 9 possible   2-path relations.  

While such an exploration sounds inviting, an important caveat is required here.  Each of these  signatures represents a different theoretical claim/prediction, but it is not the case that each is empirically distinguishable from all other predictions at the aggregated network level.  For example, consider the balanced triad signature \textit{<n.np>},    Ego sends a negative tie to Alter and a negative tie to X; Alter sends  a positive tie  to X.  The prediction represented by this signature is that, conditioned on Ego disliking a third party (X) and Alter likes that same third party,  this promotes Ego to send a negative tie to Alter.  In assessing this statement, the correlation takes into account how many of these particular conditional 2-paths  there are between Ego and Alter.   

Now consider a second configuration/prediction, also balanced, represented by the triad signature: \textit{<n.pn>}.  Here the primary difference is that the conditional 2-path represents Ego as having  a positive relationship with X and Alter has a negative relationship with X (rather than vice versa). While the balance prediction is the same, the theoretical process is different.  Indeed, in the first \textit{<n.pn>} case, at least Ego has one positive tie, which she may wish to keep.  In the second case, Ego is surrounded by negative ties, and perhaps is less committed to maintaining them.  

These two predictions, even though both represent balanced states, are quite different from one another on the surface. Yet, while the theoretical mechanism for balance and adjustment might be arguably dissimilar in these two cases, a closer look reveals that they are empirically indistinguishable at the network level.  

Consider the particular triple in the first case, \textit{<n.np>}, where Ego sends a negative tie to Alter.  We found in that particular Ego\,\text{--}\,Alter pair that there existed a certain number of 2-paths of the form \textit{<nz>}. Let's call that number $k_{nz}$.  It is also the case that for every Ego\,\text{--}\,Alter pair where the relationship is \textit{<n>} (negative), there is exactly one corresponding pair where Ego and Alter switch roles as Ego becomes Alter and Alter becomes Ego.  For each original Ego\,\text{--}\,Alter pair, the mirrored Alter-Ego pair will also have the same set of $k_{nz}$ 2-paths, except when they switch roles the \textit{<nz>} 2-paths become \textit{<zn>} 2-paths.  Thus, $k_{nz}$ for the original Ego\,\text{--}\,Alter pair will equal $k_{zn}$ for the mirrored Alter-Ego pair.  This equality will repeat for each original Ego\,\text{--}\,Alter pair. These equivalent and commensurate frequency counts of dyads and 2-paths will be repeated for every Ego\,\text{--}\,Alter pair in the multigraph.  The result is that the correlation between \textbf{N} and \textbf{NP} will equal the correlation between \textbf{N} and \textbf{PN.}   They will  necessarily be identical, even though they theoretically describe two different mechanisms.  

We claim that these two configurations, these two triad signatures \textit{<n.pn>} and \textit{<n.np>} are behaviorally indistinguishable, or \textit{phenotypical twins} , in that no evidence can be observed in any network that would allow one to know which of the two theoretical mechanisms is a better explanation of the tendencies in the network toward tie formation or maintenance.  While their genotypes are different (they have different fundamental signatures and theoretical underpinnings), their observable  manifestations and influences, at least as indicated by exact correlations between the predicted relation and the conditional 2-paths, are not.

Not all signatures have phenotypical twins (hereafter referred to simply as ``twins'').  Indeed, of the 27 signature triads, nine have no twin.  But the remaining 18 do have exactly one twin.  It is important to know which signatures have a twin, and what the twin's signature is, because it will be impossible to tell which theoretical explanation deserves credit for the observed strength in predictive power as indicated by the balance correlation. It will call for a mandatory dose of humility on all our parts.

We provide the following theorem for uncovering twins in triad signatures (a general proof of this theorem on multi-digraphs is provided in S1 Appendix.):

\begin{theorem}
\label{theorem.1}
    For any triad signature \textit{<q.rs>}, if it has a twin, it will consist of the signature \textit{<q.sr>}.  If and only if s=r (that is, s and r are the same relation), then \textit{<q.rs>} will not have a twin.
\end{theorem}

Thus, \textit{<n.pn>} and \textit{<n.np>} are twins.  Note, however, that \textit{<n.pn>} and \textit{<p.nn>} are not twins, even though they are isomorphic in their graph structures. It is important to note that non-twin signatures can, in any given graph, have identical balance correlations.  There may be several balance correlations equal to $0$, for example, indicating that there is no tendency toward balance or unbalance as predicted in those signatures.  The critical issue with twin signatures is that their Balance correlations will be equal \textit{by construction} in any and all possible graphs of any size composed of any arrangement of \textit{<p>}, \textit{<n>} and \textit{<z>} ties.
As we will see below in our analysis, isomorphic triads but which are not twin signatures, can have dramatically different balance correlations.  Indeed, in some cases, we will observe isomorphic triadic signatures that will reveal opposite signs; that is, one signature will exhibit balance while its isomorphic counterpart will be unbalanced.

\section{Empirical example}
\subsection{Background}
The importance of Balance Theory for international relations has been widely recognized \cite{moore_structural_1979, antal_social_2006, doreian_structural_2015, kinne_local_2023, zhang_risky_2023}. Current models of (structural) balance theory allow for consideration of either two types of relations (positive, or negative) (see \cite{harary_measurement_1959, aref_balance_2019}). Any further refinement in strength, or nominal categorization of relationships is impossible within the current framework. The proposed approach in this paper does allow to investigate many more refinements, which occur in international relations. Note however, that although we are studying data over time, this is not a study into the dynamics of international relations. Rather we identify balance correlations per year that subsequently substantiate a time series. In other words we study embedded triadic balance behaviour at a system level, not the feedback of system level changes into the considerations of countries when making international relational choices, although this is a further extension we would highly support, it is not studied in this paper.

\subsection{Data}
To demonstrate the usefulness and flexibility of this model, we will use the Correlates of War (CoW) data.  These data are the result of an ambitious project started in 1969 by a political scientist at the University of Michigan, David Singer \cite{singer1988reconstructing, bremer2022correlates} (for a current and exhaustive review). 
This data set includes a comprehensive accounting of both formal defense alliances and military acts of aggression that occurred each year between all pairs of countries from 1816 to 2007.  We considered any formal defense alliance between the pair of countries as a positive tie.  Acts of aggression recorded in the data set included many categories, ranging from threats of military action against another country to all out wars or invasions (see Table~\ref{table:CodesNegativeTiesNations} taken from \cite{palmeri2020}).  We coded any aggressive act by one country against another as a negative tie. In the rare case where both a defense alliance and an act of aggression occurred within the same year, we coded that tie as a negative tie for that year.

As detailed in \cite{palmeri2020}, there were 22 CoW codes regarding negative ties between nations.  They clustered these categories into five buckets, each representing a  level of hostility between nations, where 1 represented no military action and 5 represented a full scale war.  Instances of no military action  (level 1 per their coding scheme) were not coded as a negative tie in our analysis. The remaining 21 codes, however, were deemed a negative tie in this paper.  See Table~\ref{table:CodesNegativeTiesNations} for a full listing of these codes.
%In a separate CoW document, intra-state wars are covered. We did not include intra-state wars in our data.

\begin{table}
\centering
\caption{{CoW codes for negative ties between nations.} }
\begin{tabular}{llll}
\toprule
Code $2$ & Code $3$ & Code $4$ & Code $5$\\
\cmidrule{1-1} \cmidrule{2-2} \cmidrule{3-3} \cmidrule{4-4}
Threat to use force & Show of force & Occupation of territory & Begin interstate war\\
Threat to blockade & Alert & Seizure & Join interstate war\\
Threat to occupy territory & Nuclear alert & Attack &\\
Threat to declare war & Mobilization & Clash &\\
Threat to use CBR weapons & Fortify border & Declaration of war &\\
Threat to join war & Border violation & Use of CBR weapons &\\
& &Blockade &    \\

\bottomrule

\end{tabular}
\begin{flushleft}
\textbf{Note:} The above 21 coding categories used in the CoW data are organized into severity buckets, labeled $2$ through $5$, where $5$ is the most severe type of event (see \cite{palmeri2020} for details).  We left out the category `No militarized action' Code $1$ as this did not reasonably indicate a negative tie.   
\end{flushleft}
\label{table:CodesNegativeTiesNations}
\end{table}

Under the assumption of limited relational resources it is reasonable to argue that countries will consider relations with the whole set of other countries. They are limited in developing ties due to resources restrictions, which differ per country. It is recognized that the implicit or explicit neutrality choice is an embedded triad choice driven by a set of possible motives including geographical distance, but render this a political decision.

\subsection{Testing balance correlations}
Testing balance correlation raises the question: against which null distribution? We rely on a Monte Carlo-simulation to generate a sample of random networks from a data generating process (DGP) assumed for the population. This DGP preserves the  marginal distribution of negative, positive and zero ties observed in the data for each year.  For example, if a country has a defense alliance with 12 other countries, and engages in an act of aggression with 5 other countries, then the DGP would generate sample networks that would preserve this number of positive and negative ties with other countries in expectation.

To construct such a DGP, we make use of the result that the maximum-entropy model with specific expectations for some functions of the random variable of interest is (to use modern statistical terminology) an exponential family model. Its sufficient statistic must comprise the functions whose expectations are to be fixed \cite{jaynes_information_1957}. Exponential-family random graph models (ERGMs) have been developed both for scenarios in which positive and negative ties may co-occur and must thus be modeled jointly \cite{huitsing_univariate_2012} and for scenarios in which they cannot \cite{fritz_exponential_2022}, with the general polytomous relationship case first specified by \cite{robins_logit_1999}. Our application then requires the sufficient statistic vector to be the concatenation of positive and negative tie degrees of each node, resulting in a model of the form
\begin{equation}
P(\mathcal{G};\boldsymbol{\eta}) = \exp({\boldsymbol{\eta} \cdot [\textbf{1}_n^{\top}\textbf{P}, \textbf{1}_n^{\top}\textbf{N}]^{\top}})/{\kappa(\boldsymbol{\eta})},\label{eq:ergm}
\end{equation}
where $\textbf{1}_n$ is an $n$-vector of 1s and $\kappa(\boldsymbol{\eta})$ is the normalising constant so that $P(\mathcal{G};\boldsymbol{\eta})$ sums to 1 over the sample space of signed networks of interest. Estimation and simulation are then possible using general techniques for valued ERGMs \cite{krivitsky_exponential-family_2012}, or by a specialized implementation. In a parallel development, this and other null models for signed graphs were studied by \cite{gallo_testing_2024}, including an optimized implementation.

In addition, this null-generating process all preserves the observed covariance in distribution of negative, positive and zero ties in each year. 
This constitutes a conservative test, one that restricts the sampling space support taking the empirical marginals of the data matrices as expected degree. This degree based approach is more conservative than for example an Erd\"{o}s-Renyi random network approach \cite{gallo_testing_2024}, because the latter would allow a wider range of possible random networks to be considered. In our data set of country ties, we assume that there is diversity between countries in their capacity to engage in positive and negative ties. Note that this doesn't imply countries with less capacity are not making active choices. Their choice sets differ, but we assume they can and do consider all possible ties explicitly or implicitly. This assumption supports our choice to base the test on degree expectations, as it includes the investment distribution all countries make.

\begin{figure}%[p]
        \caption{{Example of balance correlations under the null distribution}}
    \includegraphics[width=\textwidth, page=1]{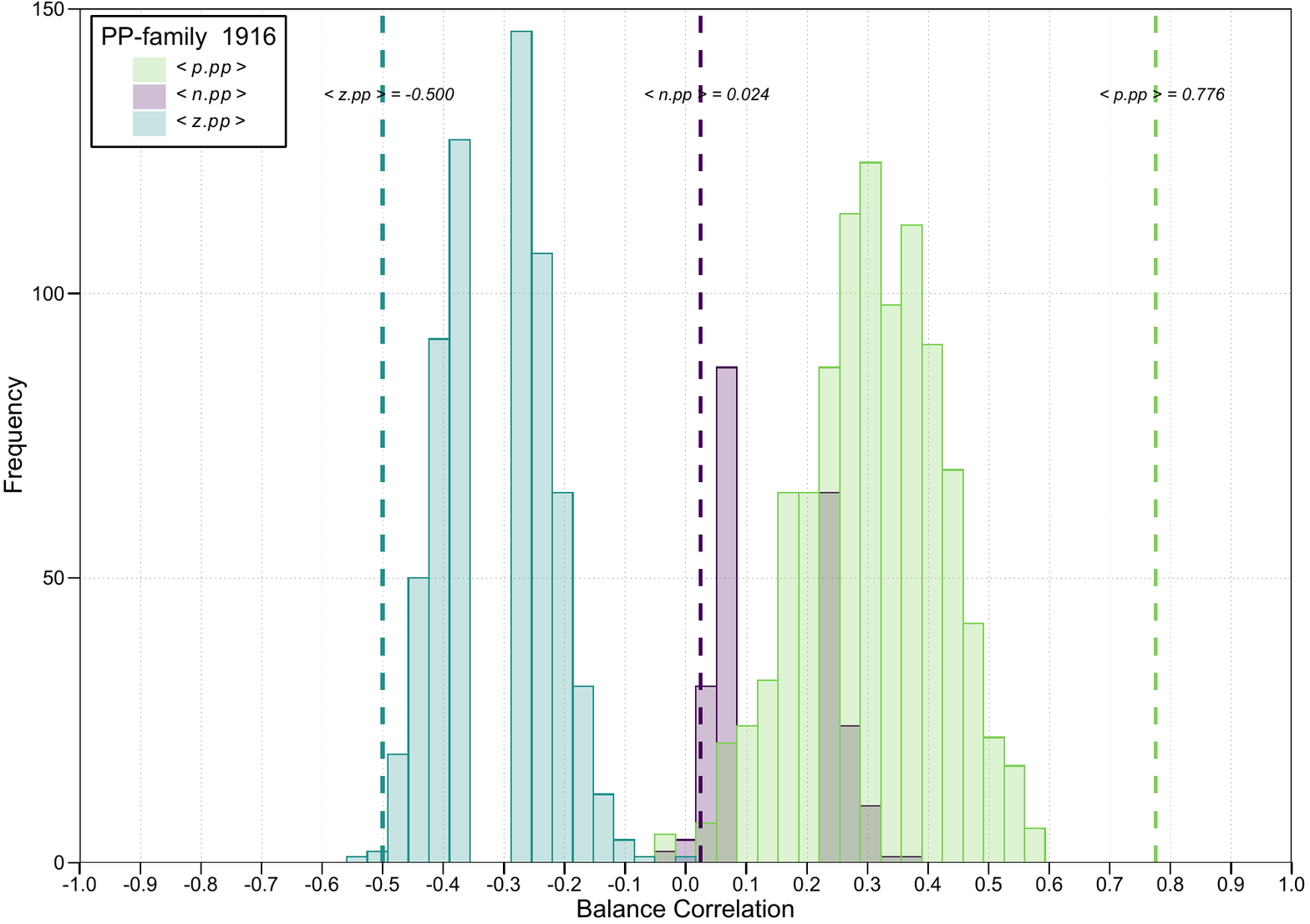}
    \centering
        \label{fig:null.1916}
\begin{flushleft}
  \textbf{Note: } In this example, $1000$ graphs were generated from a null-model that simulated networks that had expected \textit{p}, \textit{n}, and \textit{z} degrees equal to those observed in 1916. The observed densities of the positive, negative and zero ties were $.023$, $.036$, and $.940$, respectively. This histogram displays the distribution of three balance correlations (\textit{<p.pp>}, \textit{<n.pp>}, and \textit{<z.pp>}) calculated from these null graphs.  The green-shaded part of the histogram, mostly on the right side of the chart, displays the frequency distribution of the \textit{<p.pp>} correlations.  The teal-shaded part, mostly on the left side of the chart, displays the frequency distribution of the \textit{<z.pp>} correlations.  The purple-shaded portion of the histogram, in the middle of the chart, displays the frequency distribution of the \textit{<n.pp>} correlations.  The dashed vertical colored lines represent the corresponding  balance correlations for the observed data for that year.  The observed balance correlations for both \textit{<p.pp>} (the vertical green line) and \textit{<z.pp>}  (the vertical teal line) were outside the null region, suggesting they were statistically significant coefficients.  The observed  balance correlation \textit{<n.pp>} (the vertical purple line), however, was within the null distribution region and therefore not statistically significant.   
\end{flushleft}
\end{figure}

This DGP approach is based on the observed data for each given year. Thus, the set of probable null graphs can be seriously constrained, and these constraints can change for each balance correlation for each year.   The resulting distribution of null graphs may exhibit unfamiliar shapes.  For example, in Fig~\ref{fig:null.1916} we see the distribution of balance correlations for the set of \textit{<pp>} 2-path balance correlations (\textit{<p.pp>}, \textit{<n.pp>}, and  \textit{<z.pp>}).  The  \textit{<p.pp>} balance correlations are mostly positive, ranging from just below zero to $.60$.  The  \textit{<z.pp>} balance correlations, however, are mostly negative, ranging from $-.55$ to $.00$  The distribution for the \textit{<n.pp>} balance correlations, ranges from $-.05$ but a range from just below that up to just below $+.40$.  The shape of each of these distributions depends on the number of countries in the data set that year, the  density of each type of tie, the degree distributions of each type of tie, and the covariances among these degree distributions. Especially for the latter two distributions this leads to stron bi-modal distributions as can be seen from Fig \ref{fig:null.1916}. These peculiarities may differ per year. Therefore a separate DGP was used to test each balance correlation for each year.

\subsection{Results}

Before we present the primary balance results,  it is important to comment on the context for these data.  Overt acts of aggression (our source of negative ties) vary over the years for which we had data (1816 to 2007).  In a few of the earlier years, there were no acts of aggression at all.  For such years, we do not, indeed cannot, compute a balance correlation.  Also, the number of affirmative defense agreements between nations varies considerably over this period of history.  And, of course, this means the proportion of zero ties will also vary over this time.
\begin{figure}
        \caption{{Log of number of nodes, dyads, and $p$, $n$ and $z$ ties per year}}
        %\vspace{-.7in}
    \includegraphics[width=\textwidth]{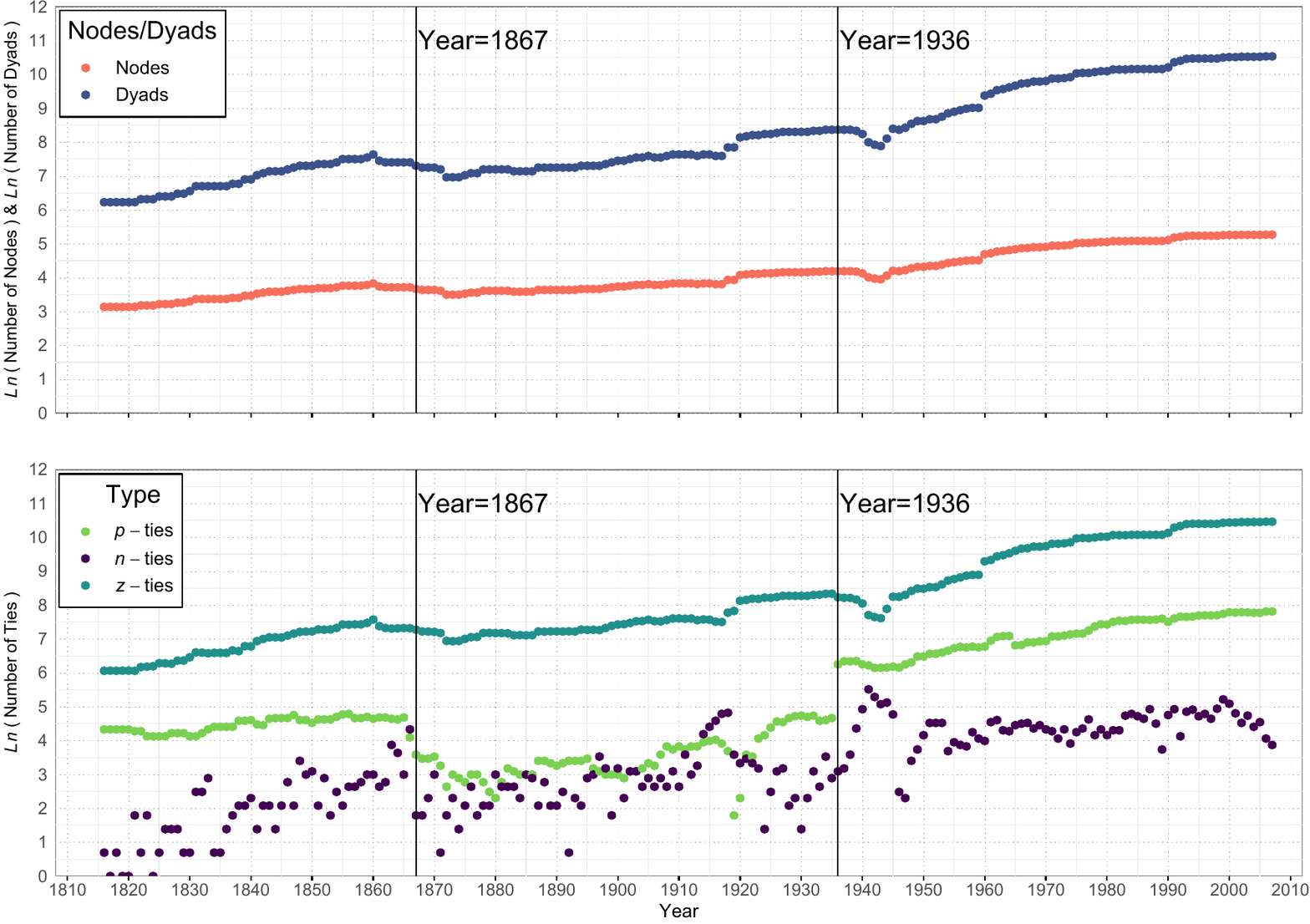}
    \label{fig:density}
\end{figure}

\begin{figure}
        \caption{{Density of $p$, $n$ and $z$ ties per year}\label{fig:density2}}
        %\vspace{-.7in}
    %\includegraphics[width=\textwidth]{Density of ties per year v19.4.1.a.pdf}
        \includegraphics[width=\textwidth]{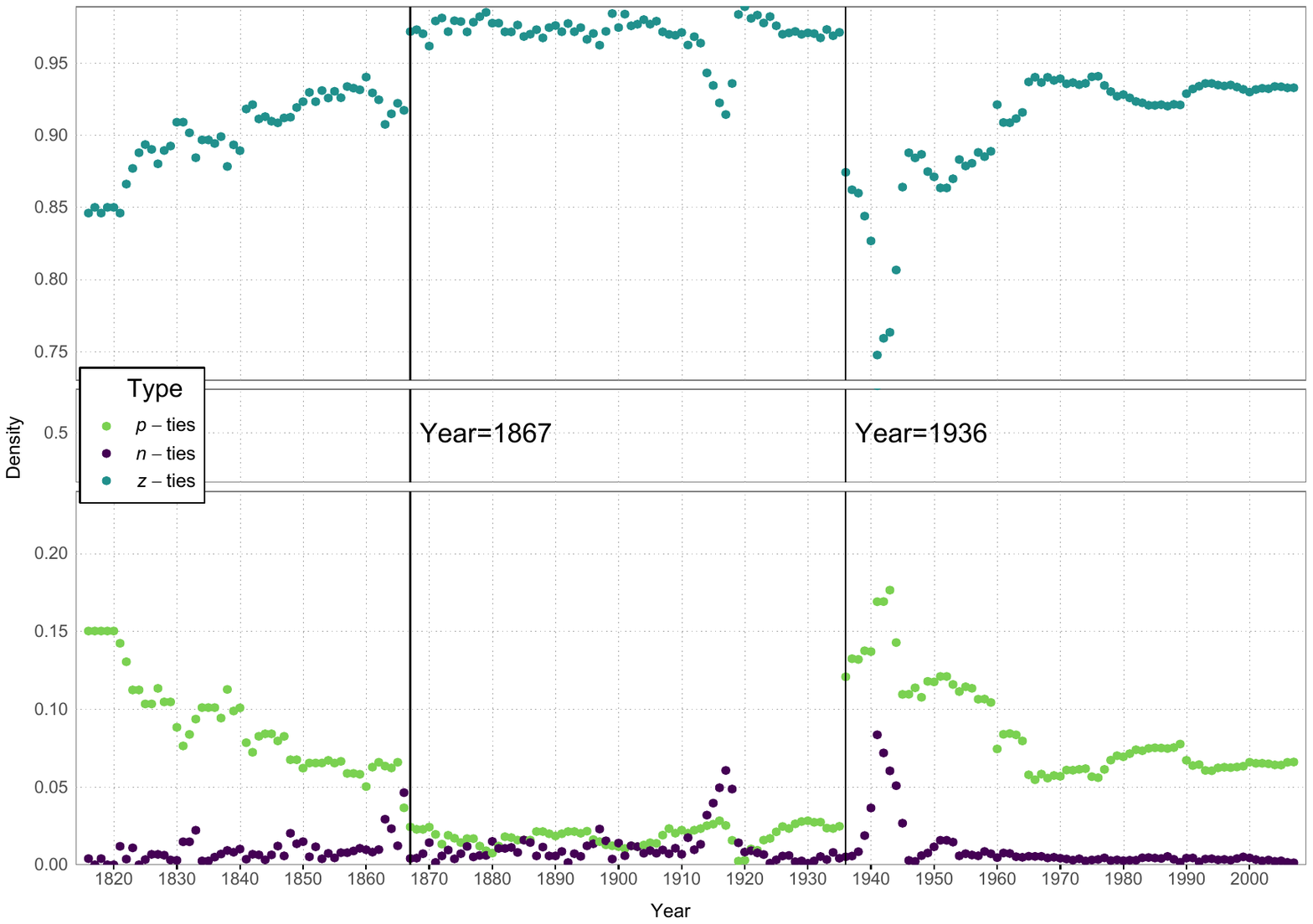}
\begin{flushleft}
  \textbf{Note:} The three dotted plots show how the density of the positive, negative and zero ties varied over the 192 years studied.  The green dots indicate the density of positive ties, the purple dots indicate the density of negative ties, and the teal dots indicate the density of the zero ties.  These densities within any given year by definition sum to 1.0. Of special note is the period of time between 1867 and 1936, the LOA years, during which there was a precipitous decline in the density of positive ties (alliances) between nations.   
\end{flushleft}
\end{figure}

In Fig~\ref{fig:density2}, we plot the density of ties of each of the three types over this time period.  First and foremost, it is clear that Z-ties are the most prevalent, with densities ranging from .75 to .99 over these 192 years.  Negative ties are usually (but not always) less frequent than positive ties, with spikes in density most notably during WWI and WWII.  Perhaps most interesting is the pattern of positive ties.  In general, the density of international relations characterized by formal defense alliances of some type varied between almost none (1909) to almost 20\% (at the conclusion of WWII).  What is striking in this plot is that these alliances were most frequent before 1867 and after 1936. In between these two years, the density of inter-nation defense agreements dropped considerably.  1867 was the beginning of the dissolution of the Austro-Hungarian empire; 1936 was the onset of WWII with the re-militarization of the Rhineland. We will call this period the  ``Lack of Alliances'' (or LOA) period.  We use it here to refer to this specific span of years (1867 to 1936), one  that we found to be critical in our exploration of balance, given the regularity with which patterns of balance were interrupted during this period.  It can be characterized as a period not of relative calm as much as a period of disorder, where alliances were either not sought or not lasting. 

We organize the presentation of the balance correlations into \textit{families}.  A \textit{family} is defined by the 2-path conditional.  For example, the \textit{<pn>} family consists of the triad signatures \textit{<p.pn>}, \textit{<n.pn>}, and \textit{<z.pn>}. 

Thus, we present three results for each of six families (\textit{<pp>}, \textit{<pn>}, \textit{<pz>}, \textit{<nn>}, \textit{<nz>}, and \textit{<zz>}) below.
We leave out  the  phenotypical twins (\textit{<np>}, \textit{<zp>} and \textit{<zn>}) from our presentation of the results below since they would be redundant with \textit{<pn>}, \textit{<pz>} and \textit{<nz>}, respectively.  

In each of the following figures, we plot the balance correlations for each year  across the 192 years for which we have data (1816 to 2007).  In addition, we show the 95\% interval from the DGP simulated null distribution in gray. 
 Thus, points representing balance correlations outside that gray region can  be considered statistically significant in that they would not be expected to appear under the null model. Balance correlations that are within the bounds of the null model, thus are not deemed statistically significant.

\begin{figure}%[p]
        \caption{{PP family of balance correlations}}
        \includegraphics[width=\textwidth]{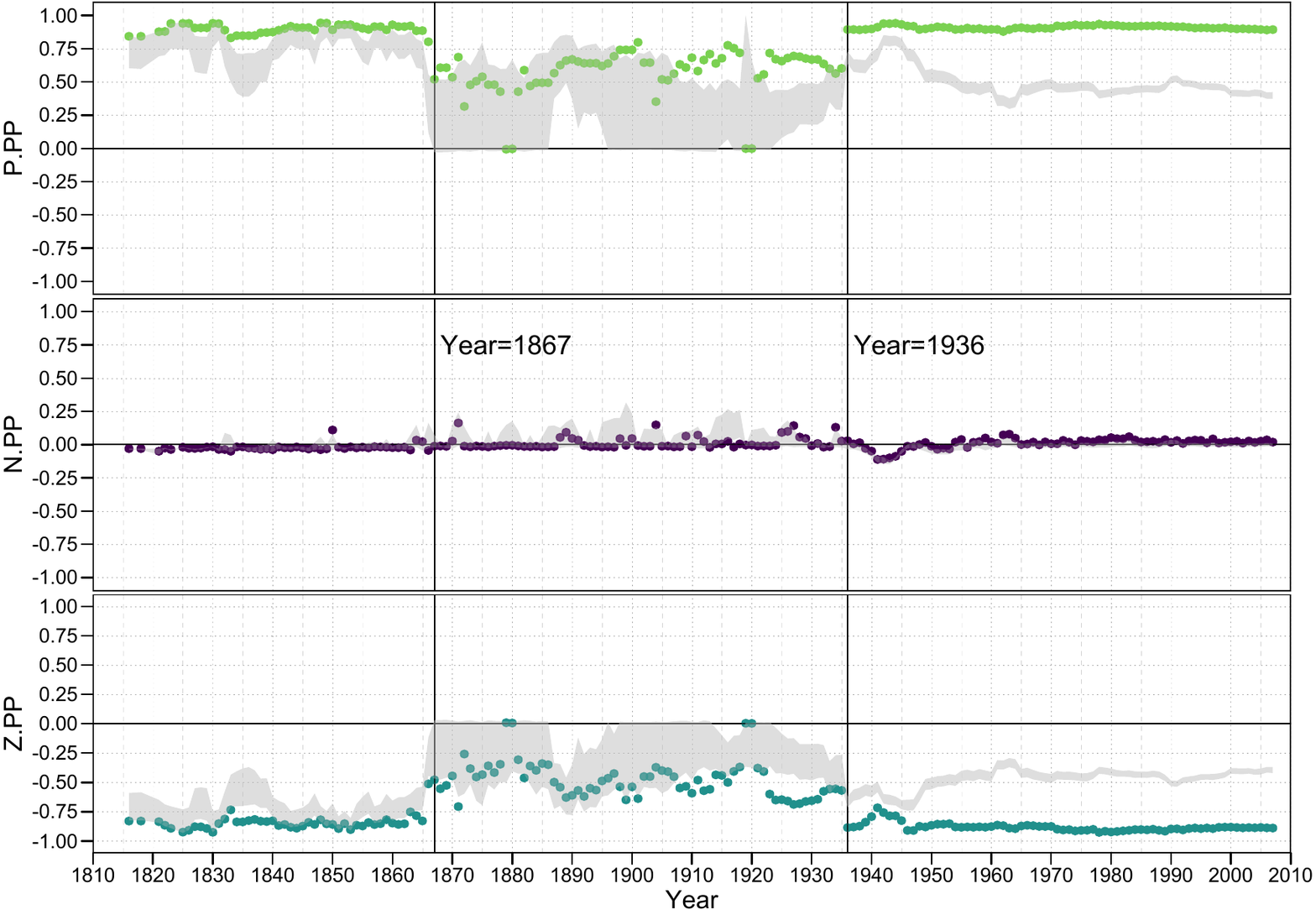}
        \centering
        \label{fig:PP}
\end{figure}

%\newpage
\subsubsection{PP family}

Fig~(\ref{fig:PP}) contains the three plots of the \textit{<pp>} family of balance correlations for \textit{<p.pp>}, \textit{<n.pp>}, and \textit{<z.pp>}.  In the first plot, we examine the pattern of the classic balance claim that a ``friend of a friend is a friend.''  We see that the \textit{<p.pp>} balance correlations For the most part are persistently strong and significant, ranging mostly from .8 to .95, providing strong evidence for this balance condition.  The primary exception to this lies within the  LOA period we identified earlier.  During this time, 
we can see that the country level behavior in the world tends to result in less balance on a global level, while it can also not be inferred that the observed balance structure is due to balance oriented behavior. It is clear, also, that the low number of positive alliances gives rise to a wider band of balance correlations under the null model.  This suggests system wide turmoil.

Remarkably, there is much less evidence for a persistent pattern of unbalance represented by the \textit{<n.pp>} triad signature. According the classic balance theory, a friend of a friend should \textit{ not } be an enemy, or at least is less likely to be an enemy.  If this principle were true, then this should yield a strongly negative correlation.  Indeed, there is  little evidence that this condition lowers the probability of a negative tie between the two nations; these correlations hover around zero for almost the entire observed period.  

While negative ties are not suppressed in the face of 2-path alliance relations, zero ties are clearly so.  The null distribution (the gray region in the plot) suggests that the structure of the distribution of positive, negative and zero ties makes a randomly generated balance correlation more likely to be negative, it is still the case that zero ties are much less likely to appear than can be attributed to chance under this null model.  Again, the primary exception to this general finding occurs during the  LOA period.  It appears for the \textit{<pp>} family, countries are likely to form alliances with other countries where they have many common mutual alliances with third party countries.  And, while they may or may not engage in a hostile stance against these countries, they appear to avoid a neutral or non-tie relation with them.  That is, they are forced to either fish or cut bait, to choose sides.

\begin{figure}%[p]
        \caption{{PN (and twin NP) family of balance correlations}}
    \includegraphics[width=\textwidth, page=1]{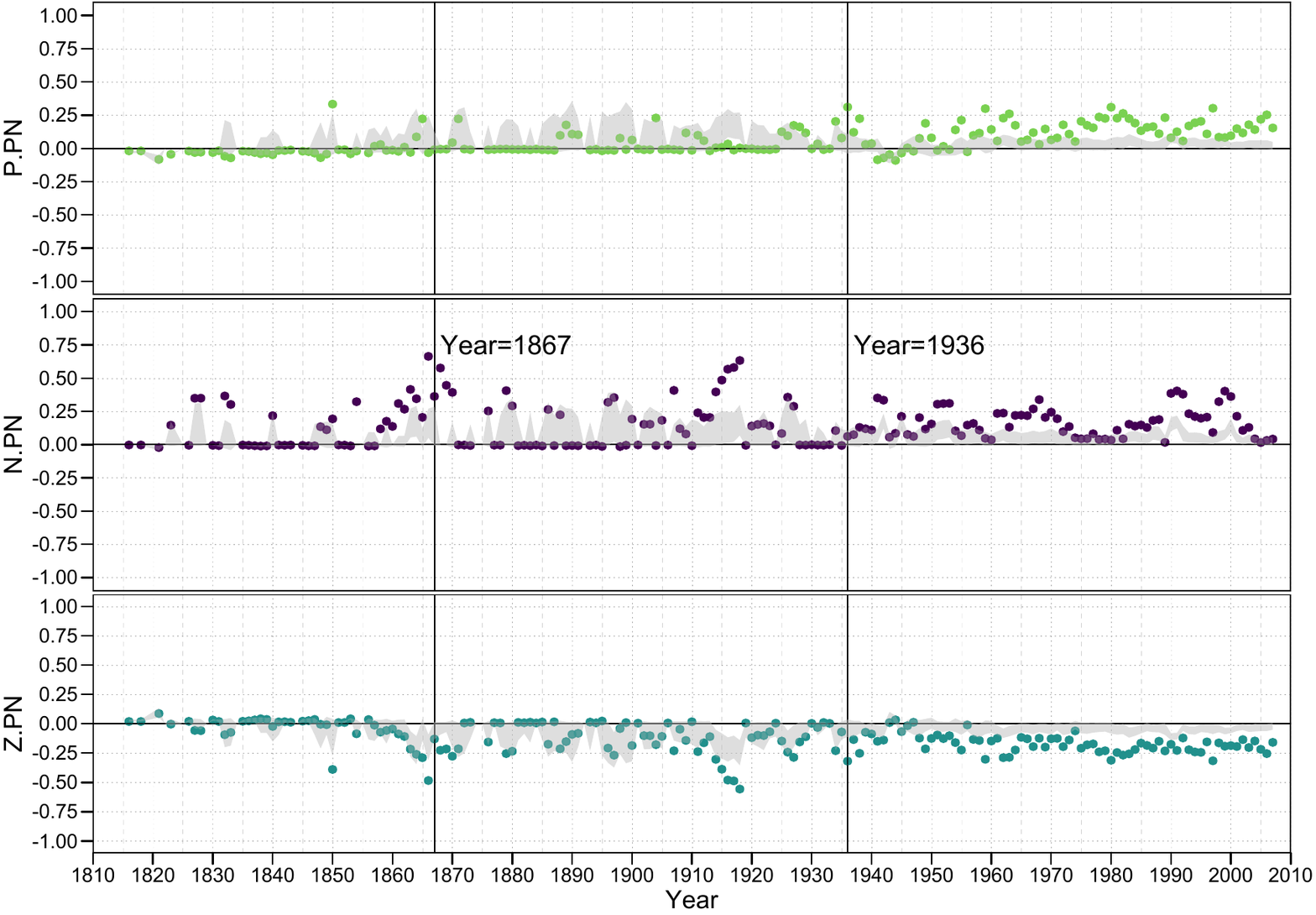}
    \centering
    \label{fig:PN}
\end{figure}

\subsubsection{PN (and twin NP) family}

Fig~\ref{fig:PN}, showing the three plots of balance correlations in the PN family, provides an interesting extension of the results in the prior figure.  While the balance correlations for \textit{<p.pn>}, the first panel in  Fig~\ref{fig:PN}, show a predominent tendency towards correlations near zero up to the years just prior to WWII, the period subsequent to WWII contain mostly significantly positive correlations.  This is true despite the fact that \textit{<p.pn>} configuration theoretically represents an unbalanced condition: One should not consider an enemy of a friend (or its twin condition, a friend of an enemy) a friend.  Yet it seems to be a consistent strategy since the late 1950's. Establishing these structures is a tertius gaudens strategy \cite{Burt1992} .  

On the other hand, the second plot in Fig~\ref{fig:PN} shows evidence for balance in that for a majority of years the \textit{<n.pn>} configuration is also positive, significant and never negative over this period. And the strength of these findings, as represented by the size of the correlations, exceeds the size of the balance-disconfirming correlations in the \textit{<p.pn>} plot.  That is, during much of this period of time we examined, the propensity to engage in a hostile act against a country who is a friend of an enemy is substantially greater than if they are not a friend of an enemy.  The fact that both of these conditions (\textit{<p.pn>} and \textit{<n.pn>}) exist at the same time speaks to the complexity of international relations.  To presume that only one balance rule operates in any given temporal or network circumstance is belied by these results.

What is also clear is the \textit{z}-ties tend to be significantly suppressed in the face of  \textit{<pn>} 2-path relations. Again, it suggests neutrality is less an option many times in the face of these 2-path conditions.   This general tendency, though, does not always coincide with the years during which this was true for the \textit{<z.pp>} balance correlations in Fig~\ref{fig:PP}.

\begin{figure}%[p]
        \caption{{PZ (and twin ZP) family of balance correlations}}
    \includegraphics[width=\textwidth, page=1]{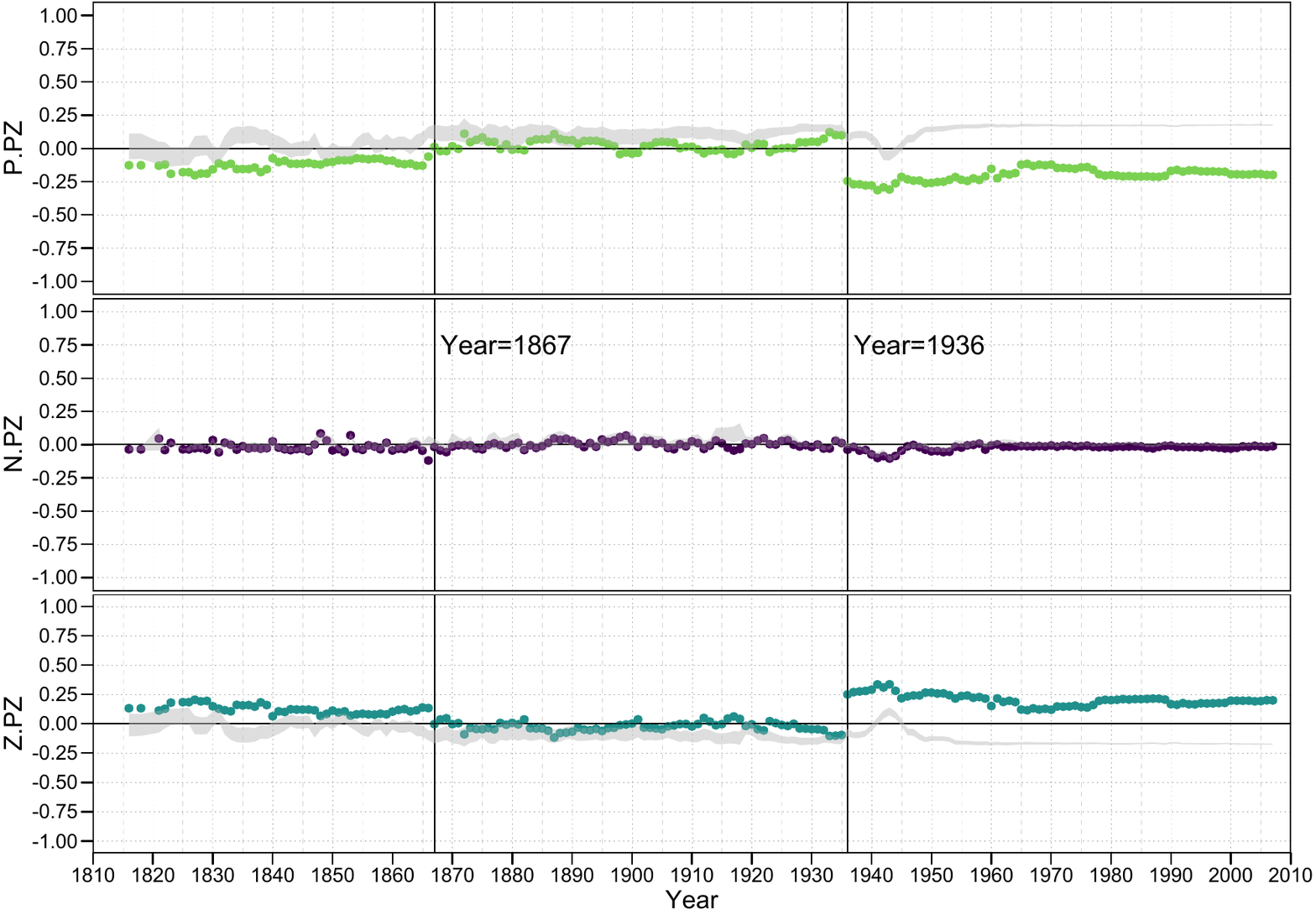}
    \centering
        \label{fig:PZ}
\end{figure}

\subsubsection{PZ (and twin ZP) family}

Fig~\ref{fig:PZ} is our first case where all members of the family are excluded from any of Heider's original predictions.  Thus, it is difficult to say whether any of these correlations should be \textit{theoretically} balanced.  What we do observe, though, is whether they appear to be \textit{empirically} balanced.  That is, are there any general patterns that show persistently positive (balanced) correlations, or perhaps persistently negative (unbalanced) correlations?  

The answer is, clearly, ``yes'' to both questions.  We find that alliance relations (positive ties) are less likely in the face of many \textit{<pz>} (or \textit{<zp>}) 2-path relations, suggesting that \textit{<p.pz>} is an empirically unbalanced configuration, at least outside the  LOA era.  And, by contrast we find that a neutral stance against another country is more likely in the face of many \textit{<pz>} (or \textit{<zp>}) 2-path relations, suggesting that \textit{<z.pz>} appears to be an empirically balanced configuration (again, outside the  LOA years).

\begin{figure}%[p]
        \caption{{NN family of balance correlations}}
    \includegraphics[width=\textwidth, page=1]{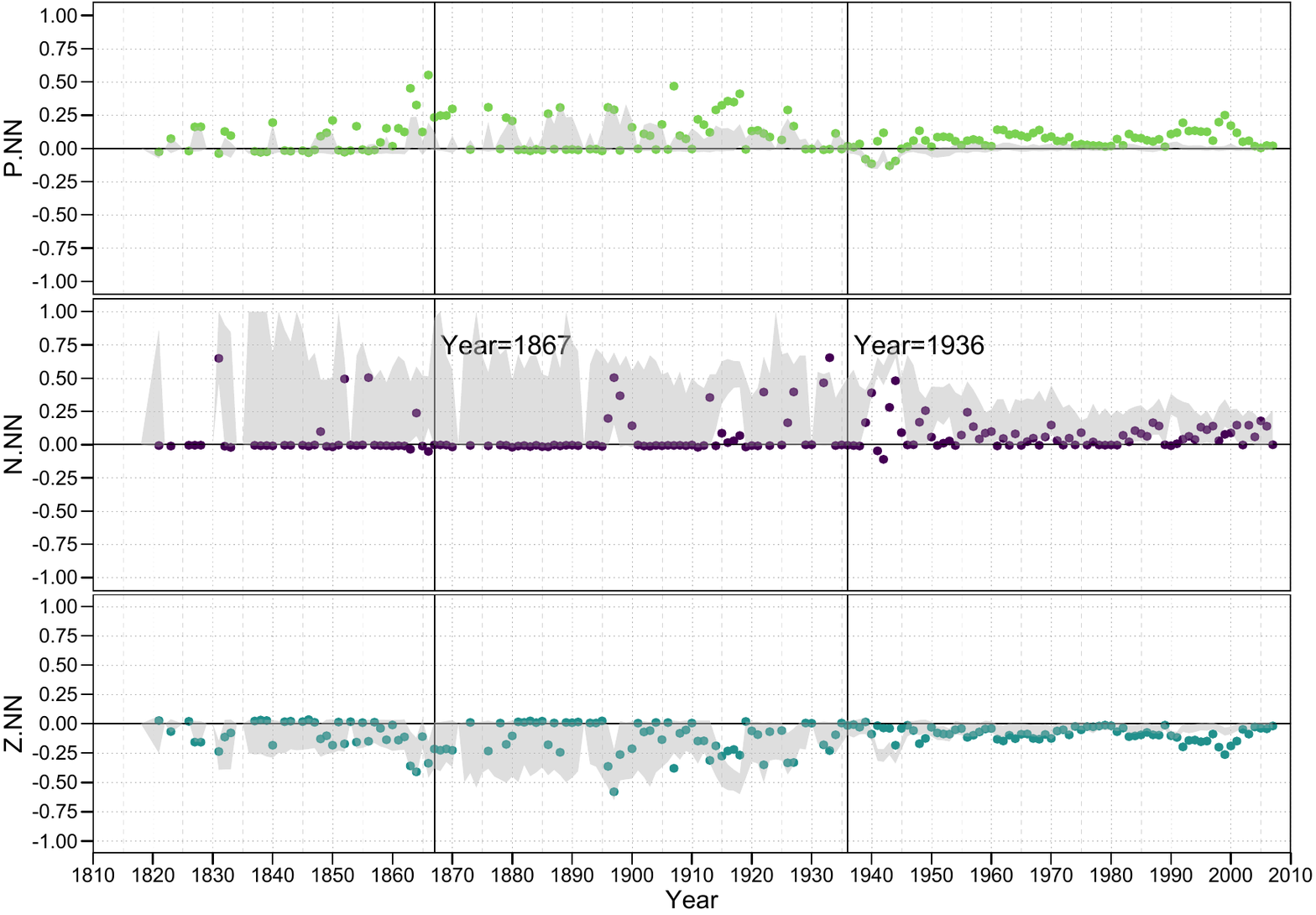}
    \centering
        \label{fig:NN}
\end{figure}
%\newpage

\subsubsection{NN family}

Fig~\ref{fig:NN} provides the balance correlations for the \textit{<nn>} family.  The first plot in this figure provides general support of the Heiderian balance proposition that ``an enemy of an enemy is a friend'' in that for over half of the  years the \textit{<p.nn>} correlation is positive and significant.  In the post LOA period, since WWII, the \textit{<z.nn>} correlation is almost always significantly negative, suggesting that neutral stances are perhaps discouraged with an enemy of an enemy.  

Of interest in the \textit{<n.nn>} plots is the large range of correlations resulting from the null model.  In particular, in the \textit{<n.nn>} plot, the null region often ranges from a correlation of zero up to .95. While most of the observed correlations are either 0 or close to it. This is largely because the density of n-ties is low to begin with, and thus the density of \textit{<nn>} 2-path ties to be even less frequent. This scarcity leaves the correlation to be quite unstable under the null hypothesis.

\begin{figure}%[p]
        \caption{{NZ (and twin ZN) family of balance correlations}}
    \includegraphics[width=\textwidth, page=1]{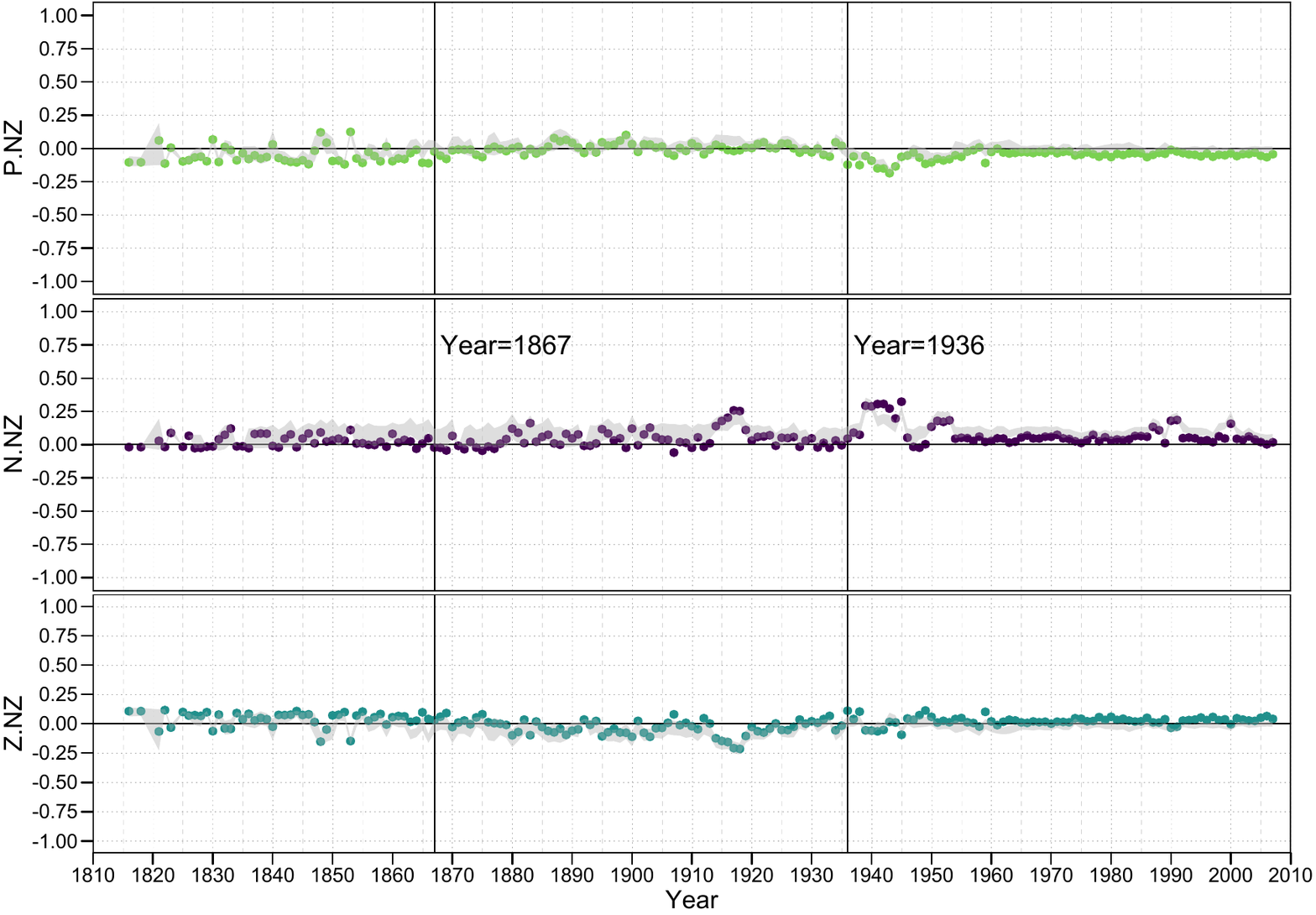}
    \centering
            \label{fig:NZ}
\end{figure}

\subsubsection{NZ (and twin ZN) family}

  Fig~\ref{fig:NZ} shows the results for the \textit{<nz>} (or \textit{<zn>}  twin) family.  Almost all of these balance correlations are either zero of close to zero, with the minor exception of the significantly positive \textit{<n.nz>} correlations during the two world wars.  Not only are the preponderance of these balance correlations near zero, they also are either  just barely border the null region or clearly not significant at all.   Thus, we have no evidence that a country who is an enemy of a neutral partner is likely to differentially engage in any particular relation (\textit{p}, \textit{n} or \textit{z}) with another nation.

\begin{figure}%[p]
        \caption{{ZZ family of balance correlations}}
    \includegraphics[width=\textwidth, page=1]{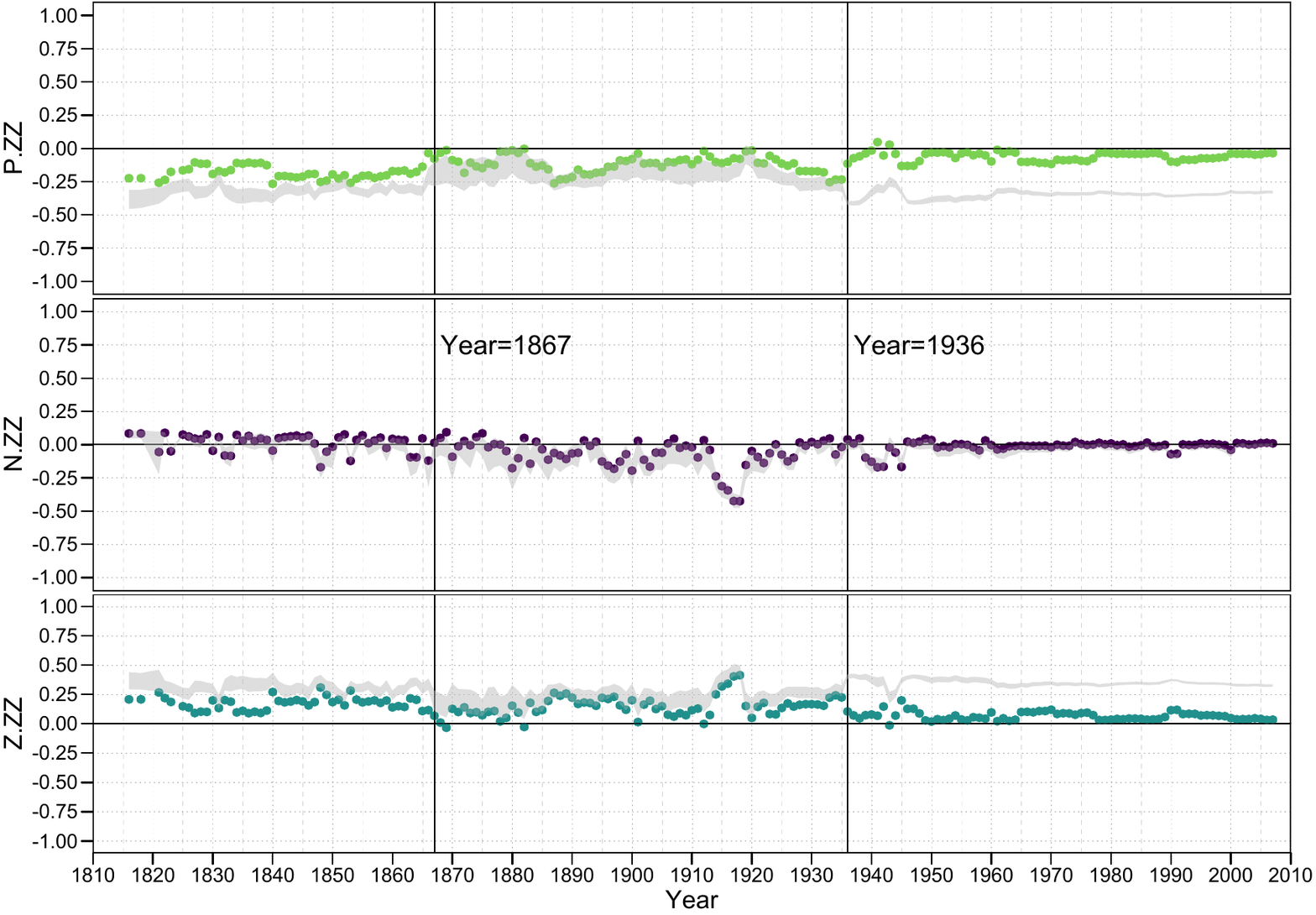}
    \centering
     \label{fig:ZZ}
\end{figure}
% \newpage

%\newpage
\subsubsection{ZZ family}

Our final set of results are provided in Fig~\ref{fig:ZZ}.  Z ties, and \textit{<zz>} 2-path ties, are the most dense.  Thus, as is displayed in the plots, the null regions are relatively thin.  The \textit{<n.zz>} balance correlations are still largely insignificant, either inside or bordering the null region.  

It is worth deliberating on the meaning of these \textit{<zz>} structures.  These suggest that the Ego\,\text{--}\,Alter dyad is surrounded by non-ties or neutral ties to an inordinate number of other countries.  It is as if Ego and Alter are in separate or separated worlds.  Thus, a \textit{<p.zz>} pattern would suggest that Ego and Alter tend to form a bridge between these relatively disconnected parties.  A positive balance correlation for \textit{<p.zz>} would indicate that an alliance is more likely if they are from these distant parts; a negative correlation would indicate that Ego and Alter are less likely to form an alliance if they are from these separated parts.  What we see from the \textit{<p.zz>} plot is that the balance correlations are generally if only slightly negative, suggesting alliances are discouraged across these chasms, but occur more frequently than expected under the null-model.  

A \textit{<z.zz>} triple tells a slightly different story.  In this case, a positive correlation would indicate that Ego\,\text{--}\,Alter pairs across such gulfs are relatively more likely to stay neutral with each other.  This would suggest that there are relatively disconnected clusters in the community, and that inside these clusters are where much of the action (both positive and negative) are happening.  Indeed, the \textit{<z.zz>} balance correlations, again outside the LOA period, are positive, if not exceptionally strong, but occur less frequently than expected. This supports the idea of a push towards global integration.

What is interesting in this family of plots, however, is that for the \textit{<p.zz>} and \textit{<z.zz>} balance correlations outside the LOA years, we find correlations that are significant (not contained in the null region) yet closer to zero than the null region itself.  This pattern is particularly striking in the post LOA years, during and following WWII.  The \textit{<p.zz>} correlations are predominantly negative and yet they are significantly \textit{greater} than we would expect by chance under the null, while the opposite holds for the \textit{<z.zz>} correlations.

\begin{figure}%[p]
        \caption{{Empirical vs theoretical Heiderian balance}}
            \centering
    \includegraphics[width=\textwidth]{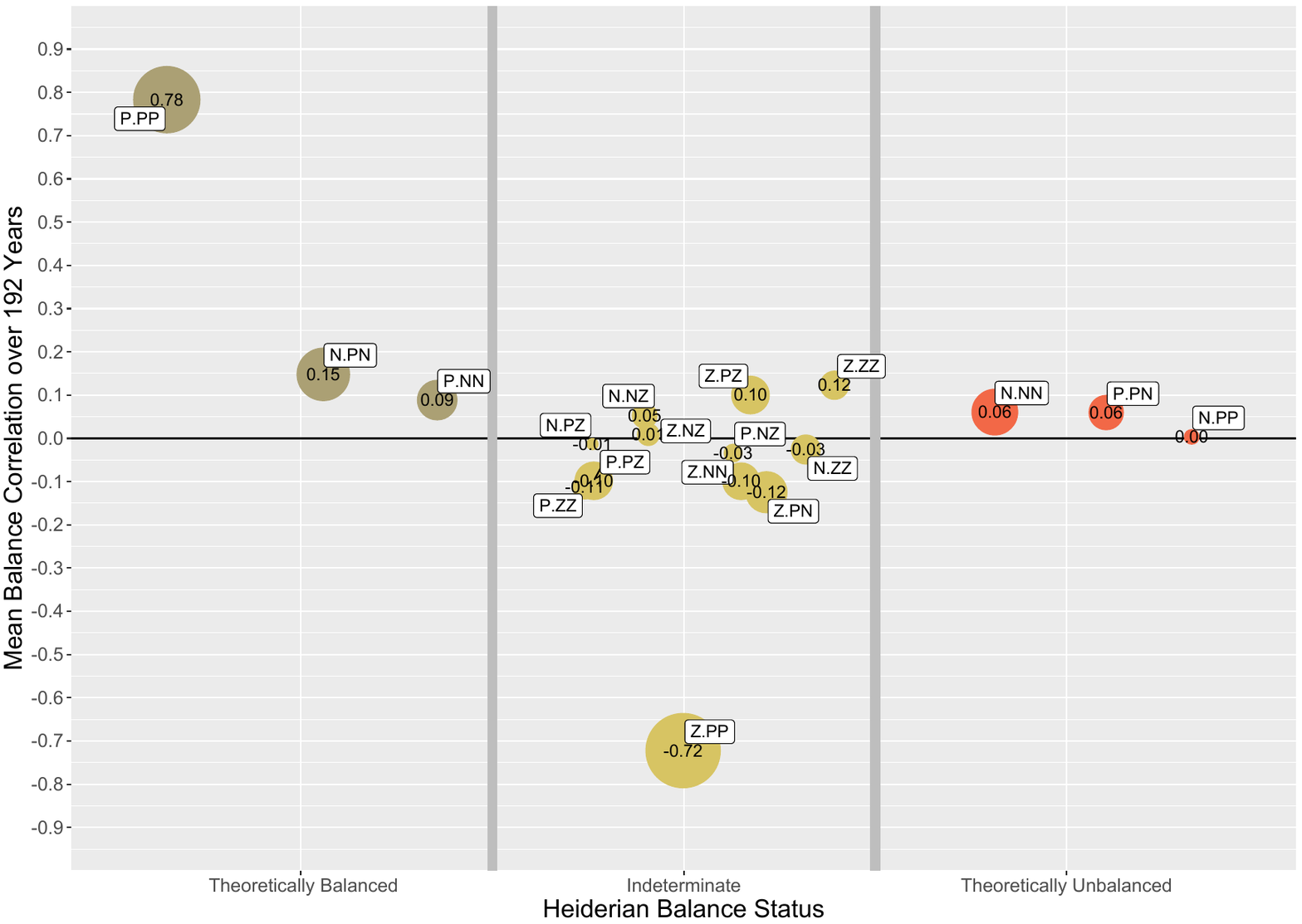}

     \label{fig:EHB}
\end{figure}

\subsubsection{Overview of results}

Fig~\ref{fig:EHB} provides a visual summary of the results across all years.  Each point in the plot represents the average balance correlation across all 192 years for which there are data.  They are divided into three groups: the ``theoretically balanced" group (\textit{<p.pp>, <n.pn>}, and \textit{<p.nn>}), the ``theoretically unbalanced" group (\textit{<n.pp>, <p.pn>}, and \textit{<n.nn>}), and the ``indeterminate" group composed of the remaining 12 triad signature configurations that contain at least one z-tie. 

Several overall findings are apparent from this figure.  First, the strongest finding supports the original Heiderian balance prediction represented by the  <p.pp> signature.  That is, at this international level, countries who have alliances with common third countries tend to have alliances with each other: `` A friend of a friend is a friend."  The other theoretically predicted balanced structures are weaker, but at least some of the <n.pn> structures are significant over time (see Fig~\ref{fig:PN}).  Interestingly, none of Heider's theoretically predicted unbalanced structures receive empirical support.  All of these correlations hover around zero. Indeed, if anything, the \textit{<n.nn>} and \textit{<p.pn>} unbalanced structures show relatively frequent positive correlations, just the opposite of what Heiderian theory would predict.

Perhaps the most interesting, and surprising, results involve those configurations with a \textit{z}-tie.  The most striking negative correlation does not emerge from any of the theoretically unbalanced triples but rather occurs in the \textit{<z.pp>} configuration.  That is, being an ally of an ally seems to carry no systematic information about hostile acts between two countries that differentiates it from other contexts. In contrast, a neutral tie is far less likely to occur in this context than in any other context. Other \textit{z}-tie results also engage in unexpected patterns, although the strength of those findings is not as evident as is the \textit{<z.pp>} case. We dilate these results and 
 based on the correlations among Balance Correlations make the case that beside the theoretical arguments for treating \textit{z}-ties as separate category, the empirical results strongly corroborate this idea.

\subsubsection{Interpreting \textit{z}-ties}

\begin{figure}%[p]
        \caption{{Correlations of balance correlations over 192 years}}
    \includegraphics[width=\textwidth, page=1]{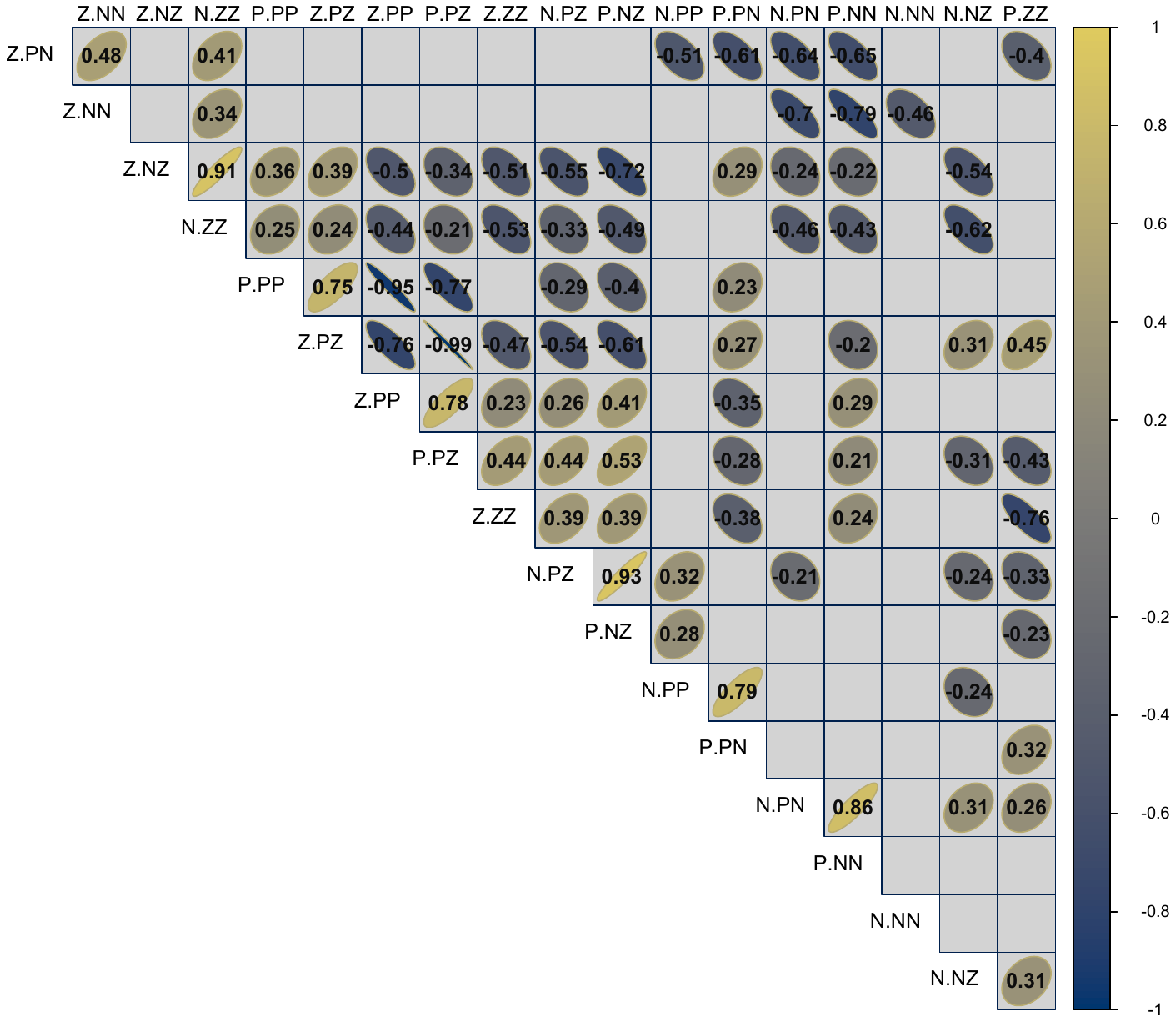}
    \centering
     \label{fig:corofcor}
\end{figure}

In our example on relations between countries, clusters of balance theoretical behaviors can be identified, i.e. clusters of balance correlations can be discerned that are positively correlated, irrespective of whether the balance correlations are significant or not. Furthermore, these clusters may show negative correlations with balance correlations in other clusters. Earlier we discussed the marking difference between \textit{<p.pp>} and \textit{<z.pp>}. Fig~\ref{fig:corofcor} shows a strong negative correlation ($-.95$) between these two signatures. So the mean values of these balance correlation are opposite, and this is consistent over time. An even stronger negative correlation occurs between \textit{<z.pz>} and \textit{<p.pz>} ($-.99$). Although the balance correlations themselves are not extreme, the observed signatures in our example seem to be highly dependent. It is stressed that this is an empirical effect not a priori determined.

Using a hierarchical clustering, looking at $7$ clusters gives a first singleton \textit{<n.nn>} (unbalanced), and $4$ pairs (\textit{<n.pn>} and \textit{<p.nn>}, both balanced; \textit{<n.pp>} and \textit{<p.pn>}, both unbalanced; and two indeterminate pairs, \textit{<n.nz>} and \textit{<p.zz>}, and, \textit{<z.pn>} and \textit{<z.nn>}). The balanced and unbalanced pairs have much stronger correlations, $.86$ and $.79$, respectively, compared to $.31$ and $.48$ for the respective indeterminate pairs. There are two larger blocks with 4 and 5 signatures, respectively. The first contains signatures from 4 different families, and includes the balanced \textit{<p.pp>}, which is most strongly correlated with \textit{<z.pz>} ($.75$), while the other two, \textit{<z.nz>} and \textit{<n.zz>} are even stronger correlated ($.91$) the other correlations range from $.24$ to $.39$. In the largest cluster all signatures are theoretically indeterminate with highest correlations between the pair \textit{<n.pz>} and \textit{<p.nz>} ($.93$), and the pair \textit{<z.pp>} and \textit{<p.pz>} ($.78$). The other correlations in this cluster range from $.23$ and $.53$.  

Now this result may helpful to assess whether the \textit{z}-ties should also empirically be considered a separate type. To interpret the valence of \textit{z}-ties as either positive or negative if we replace these \textit{z}'s with either a \textit{p} or \textit{n} designation, such that it provides a balance theoretical prediction consistent with the observed mean balance correlation (Fig~ \ref{fig:EHB}). 

\begin{table}
    % \begin{adjustwidth}{-0.2in}{0in}
    \begin{center}
    \caption{{Establishing ambiguity of \textit{z}-ties.}\label{table:ztieambiguity}}
    \begin{tabular}{b{2cm}r>{\centering}b{2.9cm}>{\centering}b{4cm}>{\raggedleft\arraybackslash}b{2.9cm}}
    \toprule
    Indeterminate signature& $\overline{BC}$ & (Un)Balanced imputed signature & Correlation with the imputed signature (CWIS) & \textit{z}-tie ambiguity? \\
    \midrule
    \textit{<z.pn>}     & $-.12$ & \textit{<\textbf{p}.pn>}& $-.61$ &Yes, CWIS\\
    \textit{<z.nn>}     & $-.10$ & \textit{<\textbf{n}.nn>}& $-.45$ &Yes, CWIS\\
    \textit{<z.nz>}     &  $.01$ & \textit{<\textbf{n}.\textbf{p}n>} or \textit{<\textbf{p}.n\textbf{n}>}& $-.24$ or $-.22$& Yes, imput. ambig.\\
    \textit{<n.zz>}     & $-.03$ & \textit{<n.\textbf{nn}>} or \textit{<n.\textbf{pp}>}& Both not sig.&Yes, CWIS\\
    \textit{<z.pz>}     &  $.10$ & \textit{<\textbf{n}.p\textbf{n}>} or \textit{<\textbf{p}.p\textbf{p}>}& Not sig. or $.75$& No, \textit{z} could be \textit{p}\\
    \textit{<z.pp>}     & $-.72$ & \textit{<\textbf{n}.pp>}& Not sig.& Yes, CWIS\\
    \textit{<p.pz>}     & $-.10$ & \textit{<p.p\textbf{n}>}& $-.35$&Yes, CWIS\\
    \textit{<z.zz>}     &  $.12$ & \textit{<\textbf{p.pp}>}& Not sig.&Yes, CWIS\\
    \textit{<n.pz>}     & $-.01$ & \textit{<n.p\textbf{p}>}& $.32$&No, \textit{z} could be \textit{p}\\
    \textit{<p.nz>}     & $-.03$ & \textit{<p.\textbf{p}n>}& Not sig.&Yes, CWIS\\
    \textit{<n.nz>}     &  $.05$ & \textit{<n.\textbf{p}n>}& Not sig.&Yes, CWIS\\
    \textit{<p.zz>}     & $-.11$& \textit{<p.\textbf{pn}>}& $.32$ &Yes, imput. ambig.\\
    \bottomrule
    \end{tabular}
    \end{center}
    
    %\begin{flushleft} 
    
        \textbf{Note:} Bold printed are $z$-imputations. In the first $z$-imputation of $n$ and $p$ in \textit{<z.nz>}, instead of \textit{<n.np>} the correlation with the twin signature \textit{<n.pn>} is reported. In imputing $p$ in \textit{<p.nz>} instead of \textit{<p.np>} the correlation with  twin signature \textit{<p.pn>} is reported. In imputing $p$ in \textit{<n.nz>} instead of \textit{<n.np>} the correlation with twin signature \textit{<n.pn>} is reported.
     %   \end{flushleft}
    % \end{adjustwidth}
\end{table}
In Table~\ref{table:ztieambiguity} little support for the idea that \textit{z}-ties can be considered either as positive or negative is found. Rather there is much support for what we call \textit{z}-tie ambiguity. When \textit{z}-labels are imputed with \textit{n} or \textit{p} labels, according to achieve a signature that is coherent with Heiderian balance theory multiple situations can be distinguished. First, in most cases coherence can be achieved by exclusive \textit{p} or \textit{n} imputation. Second, coherence can be brought about by either \textit{p} or \textit{n} imputation. Third, coherence can only be achieved by imputation of both \textit{p} and \textit{n}. 

In the latter case there is obvious ambiguity, revoking the idea that \textit{z}-ties can be considered either positive or negative. This occurs in $2$ cases ($\textit{<z.nz>}=.01$ and $\textit{<p.zz>}=-.11$), providing a first logic to consider \textit{z}-ties as a separate class on empirical grounds, besides the theoretical argument provided earlier. However, in all other cases a unique alternative labeling for \textit{z}-ties exists that will result in a coherent mean balance correlation.

In the second case where a coherent mean balance correlation occurs whenever either an \textit{n} or an \textit{p} gets imputed there is a case for which no ambiguity can be detected. In the case of \textit{<z.pz>} when \textit{z} is imputed with the \textit{p} label, a strong positive correlation ($.75$) between \textit{<z.pz>} and \textit{<p.pp>} supports the conjecture that \textit{z}-ties are to be seen as positive ties. However, in the other indeterminate signature such support is not found as correlation of \textit{<n.zz>} with both \textit{<n.pp>} as well as with \textit{<n.nn>} are not significant.

In the eight cases where one unique label creates a coherent balance correlation, only one signature has a significant correlation with the imputed signature (\textit{<n.pz>} and \textit{<n.pp>}, $\rho = .32$). In 4 cases there is no significant result, so we must conclude on ambiguity of \textit{z}-ties. Certainly not on the whole set as there are 6 undisputed signatures where \textit{z}-ties are to be treated as positive ties, and $3$ signatures as negative ties. But, neither on the context specific level as only two indeterminate signatures are consistently correlated with imputed signatures. This warrants treating \textit{z}-ties as a separate class based on empirical grounds.

\section{Extensions}

This paper deals with signed graphs, where edges between nation states are undirected symmetric ties. One of major advances in \cite{HollandLeinhardt79} was to recognize that networks are often, perhaps most often, composed of directed ties.  The work presented here can be extended to directed ties, also, leading to new specific theoretical predictions and explorations of possible balance statements and commensurate balance correlations to assess them.  

However, it should be noted that this increases substantially the number of possible balance statements.  Instead of one relation for each pair, there is both an in-tie and an out-tie that would have to be specified; thus, for each pair we must specify two relations, not just one.  For example, we could describe a case where Ego sent a positive tie to Alter, but Alter does not respond with any sentiment (i.e., send a \textit{z}-tie back).  Perhaps Ego sends a negative tie to X, and X reciprocates with a non-tie.  Further, suppose Alter sends a negative tie to X and X responds with a similar negative tie back to Alter.  This configuration would be described as \textit{<pz\,.\,nz\ nn>}.  Since there are six directed ties in any labeled triple, then the number of possible \textit{directed triad signatures} is $3^6=729$, not simply $27$ as described in this paper.  An interesting extension of our approach to digraphs is that of these $729$ signatures, $702$ have an indistinguishable twin and $27$ do not (see the proof of the theorem 
\ref{theorem.1} 
in the appendix), resulting in $378$ unique balance correlations.  We expound on these structures, their balance statements, and their empirical results in a subsequent paper on balance correlations in directed graphs.

An extension that would even further reintegrate a structural approach with Heider's original ideas is to replace an individual node as context with sets of attitudes, values, norms, or issues \cite{opp_balance_1984}. This would allow to test different versions of balance theory, and further establish the link to cognitive dissonance theory \cite{festinger_theory_1957, harmon-jones_cognitive_2019, opp_structural_2021}.

The null model we have chosen controls for both positive and negative degree heterogeneity of the nodes. The result of \cite{jaynes_information_1957} is very general, and the exponential family framework can be used to control for other social forces, such as geographic proximity and language, by augmenting or replacing the sufficient statistic in Eq~\ref{eq:ergm} with operationalizations of these effects.

\section{Discussion and conclusion}

Our intent here is to propose a new approach to the theorizing and study of balance processes in network contexts. By formally incorporating neutral or zero ties into the analysis, we extend balance theory in a significant way that better matches the real world of actors in a systemic relational structure.  And, in an important way, we return to Heider's original thinking by focusing on Ego's choice of how to relate to a specific Alter when confronted with the concomitant relations with one or many third parties.  We present a novel nomenclature, the triad signature, that assists in  identifying and theorizing about (un)balance conditions.  

Finally, we propose a simple, intuitive measure, the balance correlation, that directly reflects the extent to which Heider's predictions (and its extensions in this paper) are supported by the evidence in an observed network.  This measure captures the difference in  probabilities of Ego's choice of a tie with Alter, conditioned on whether the specific balance-theoretic 2-path configuration exists or does not exist.   We also describe a reasonable null test of this statistical measure and apply it to a well-known data set employed by political scientists in international affairs, the Correlates of War data.

As powerful as this approach is, we also note that the 27 different theoretical predictions that can be identified using the triad signature are not always uniquely identifiable empirically.  We show that, indeed, 18 of these signature have a phenotypical twin signature, in that the two signatures will necessarily receive identical support in any given network and can therefore never be distinguished using a balance correlation.  We also provide a proof that reveals the necessary and sufficient conditions for twins to exist and how to identify them. 

We further demonstrate the power of this approach by exploring the balance correlations of all  the possible predictions identified by the triad signatures over almost two centuries of conflict and alliances among extant nations in the world.  We find systematic underlying patterns that deserve attention.    First, we find strong evidence overall for balance theory on positive ties. In cases where the conditional 2-path relations contain either two  positive ties or two negative ties, the probability of a direct positive tie is significant for most of the years, as would be expected according to balance theory.

Second, we find that the predictions around unbalanced states are not confirmed for the most part and occasionally even counter to Heider's original contention. 
 For example, the theoretically unbalanced \textit{<n.pp>} correlation hovers around zero for most of the two centuries of data.  And a clear example of a result that dis-confirms an unbalanced prediction is given in Fig~\ref{fig:PN}, where the theoretically unbalanced \textit{<p.pn>} signature is consistently positive in the post WWII era. This is consistent with an alternative to balance seeking behavior, namely a tertius gaudens strategy as described in \cite{Burt1992}. However, the results also show that the balanced \textit{<n.pn>} signature is stronger. This emphasizes that both strategies may be at play at the same time in the same network, revealing the complexity of studying international relations.  

Thirdly, in a related general finding, it appears that the \textit{z}-ties (neutral ties) play a larger role in establishing balance, or at least avoiding unbalanced states, than negative hostile ties over substantial time periods.  For example, in the case of where the Ego\,\text{--}\,Alter dyad is embedded in many \textit{<pp>} 2-path relations (see Fig~\ref{fig:PP}), there is no discernible evidence that Ego avoids negative ties (\textit{<n.pp>}), but there is substantial and significant suppression or avoidance of neutral ties (\textit{<z.pp>}).  

This study of embedded triadic balance behaviour over time suggests that although the strive for balance may be a component of individual countries' motivation, it does not necessarily suggests social equilibrium or social resilience of the system. In fact, we find over time that balance can become more and less prevalent in whole networks and that these levels of prevalence can be stable over longer time spans. This is consistent with Heider's \cite{Heider1958} observation that the tension produced by unbalance may be attractive. In the setting of international relations such imbalance may hold promise for opportunities to achieve nodal benefits. The presented method indicates the prevalence of different types of balance, imbalance and undetermined behaviors.

In sum, it is our contention that balance correlations provide a simple, intuitive way to assess the extent to which specific assumption on balance behaviors might be true on the average across an entire network.  Furthermore, by incorporating agentic zeroes into the exploration of balance conditions, we may uncover new properties of social order that have been prevalent, even dominant, but  hidden from the researcher's purvue or agenda.  We hope this new approach opens up avenues of exploration that have been heretofore ignored. 

As a final remark we want to emphasize that an updated Correlates of War database (real-time) may demonstrate the practical use of balance correlation as an indicator of the level of global stability. Although we leave it to historians to corroborate that the LOA-era was characterized by dismal turbulence more than in other periods, the consistent drop in balanced relations might well serve as warning signal. If so, balance correlations may inform politicians and military leaders on global temperature in international relations.

\section*{Supporting information}

\paragraph*{S1 Appendix.}
\label{S1_Appendix}
{\bf Proof of theorem \ref{theorem.1}}
The claim in Theorem \ref{theorem.1} is based on the fact that if the indices of the two variables in the balance correlation are permuted in the same way, the correlation does not change due to associative and commutative properties of the addition of the cross-products in the correlation function. Transposing matrices is one unique permutation of the elements in a matrix.

\begin{definition}\label{matrixdef}
Let $\mathfrak{M}=\{M_1, ..., M_m\}$ be the set of $m$ square matrices. All matrices $M_q$ are non-empty with zero diagonals and are binary such that each element $M_{q_{ij}} \in \{0,1\}$. Further, $m\ge2$, and $\sum_{q=1}^m M_q = \mathfrak{L}$, is the matrix representation of the complete graph. Also, all matrices in $\mathfrak{M}$ have identical and identically ordered row and column indices, $i,j\in \{1, ..., n\}$ for the $n$ number of nodes. 
\end{definition}

\begin{definition}\label{nuggetdef}
Let  $ \mathfrak{M}^{'} =\{ M^{'}_1, ..., M^{'}_m \} $ be the set of transposes of matrices in $\mathfrak{M}$. 
A \textbf{nugget} is defined to be the element-wise product of any matrix in 
$\mathfrak{M}$ 
with any matrix in 
$\mathfrak{M}^{'}$. 
The nugget set $\mathfrak{U} = \mathfrak{M} \times \mathfrak{M}^{'} = \{U_1, ..., U_g\}$ and contains all possible $g$ nuggets. 
A nugget thus has the same indices of the matrices in $\mathfrak{M}$ and is also binary. There are $g=m^2$ possible nuggets.
\end{definition}

\begin{lemma}\label{nuggetslemma}
In the nugget set $\mathfrak{U}$ consisting of $g$ nuggets, there will be exactly $m$ symmetric nuggets, and exactly $m^{2} - m$ asymmetric nuggets. 
\end{lemma}

\begin{proof}[Proof of lemma \ref{nuggetslemma}]
Note that $M_{q_{ji}}=M_{q_{ij}}^{'}$. By contradiction, assume for any $ij$-th element in a nugget, the element-wise product of matrices, $M_{r_{ij}} \times M_{s_{ji}}=1$ and $M_{r_{ji}} \times M_{s_{ij}}=1$. Since, $M_{q_{ij}} \in \{0, 1\}$, this implies $M_{r_{ij}} = M_{s_{ji}}=1$ and $M_{r_{ji}} = M_{s_{ij}}=1$, so $M_{r_{ij}} + M_{s_{ij}}=2$. However, if $r \neq s$ this contradicts that the sum over all matrices results in a matrix $\mathfrak{L}$ (see definition \ref{matrixdef}). If and only if $r=s$ the assumption holds, as the summation visits each matrix $q$ in $\mathfrak{M}$ only once, which implies there are exactly $m$ symmetric nuggets; those of a matrix and it's transpose.
\end{proof}

\begin{definition}\label{BC}
Let a balance correlation be the Pearson correlation between elements of a nugget $U_k$
and the elements of the inner product of two nuggets $U_lU_m$, ignoring the diagonals, i.e.,
\begin{equation}\label{pearson}
    \rho(U_k, U_lU_m) = \frac{\cov(U_k, U_lU_m)}{\sqrt{\var(U_k)\var(U_lU_m)}}
\end{equation}
Now, $\mathfrak{B}$, is the set containing all \textbf{possibly unique} balance correlations that can be derived from the $g$ nuggets in any set $\mathfrak{U}$. The specific order of the nuggets (and hence the matrices they are derived from) in a balance correlation, reveal a statement about the association between a 2-path and tie in bi-directional relations. 
\end{definition}
So, $\rho(U_k, U_lU_m)$ is a quantification of the statement about an association between ties in $U_k$ and 2-paths in $U_lU_M$. This quantification differs from $\rho(U_k, U_mU_l)$ or $\rho(U_l, U_kU_m)$, which quantify statements of association between ties in $U_k$ and $U_mU_l$, and, $U_l$ and $U_kU_m$, respectively.

Now, since the bi-variate Pearson correlation is defined in terms of variances and covariance, and those metrics are permutation invariant, i.e. for any permutation function $\pi_p$ we have
\begin{equation}\label{permutation}
    \rho(U_k, U_lU_m) = \rho(\pi_p(U_k), \pi_p(U_l)\pi_p(U_m)).    
\end{equation}
Given that there are $m^2$ nuggets, and that balance correlation considers 3 nuggets, there are $m^6$ possible balance correlations. However these are not all unique, because transposition is a permutation and by definition \ref{nuggetdef}, nuggets include transposed matrices.

Nuggets, as Hadamard products that maintain associativity, imply that the order of matrices within a nugget is not important ($M_r \times  M_s^{'}=M_s^{'} \times M_r)$. However, it is relevant which matrix in a nugget is transposed in revealing the underlying theoretical statement. Furthermore, the second variable, $U_lU_m$, is a matrix inner product. It is well-known that taking the transpose of an inner product, transposes the matrices and reverses their order. There are three situations in which transposing either or both variables in $\rho$ changes the formulation of the underlying statement or in other words demonstrate equivalence between two underlying statements under all possible instantiates. Only, when both $U_k$ and $U_lU_m$ are symmetric by construction $\rho$ quantifies a \textit{sole} statement. In the other cases we speak of \textit{twin} statements.

\begin{proof}[Proof of theorem \ref{theorem.1}]
    Assume nuggets $\{U_k,U_l,U_m\} \in \mathfrak{U}$ and define $\{U_k=U_q^{'},U_l=U_r^{'},U_m=U_s^{'}\}$ which implies $\{U_q,U_r,U_s\} \in \mathfrak{U}$. Since, transposing is a permutation, by Eq~\ref{permutation} we have 
    \begin{equation}
        \rho(U_k, U_lU_m) = \rho(U_k^{'}, U_m^{'}U_l^{'}),
    \end{equation}
     which by substitution gives $\rho(U_k, U_lU_m) = \rho(U_q, U_sU_r)$. If and only if, $U_k=U_k^{'} \neq U_q$, and,  $U_lU_m = U_m^{'}U_l^{'} \neq U_sU_r$ no other quantification exists that will always produce the same quantity. By the fact that there exists one and only one possible transpose of each nugget, in all other cases there exists exactly one other configuration of distinguishable nuggets that will lead to the same quantification, where $U_k=U_k^{'}=U_q$, and,  $U_lU_m = U_m^{'}U_l^{'}=U_sU_r$.
\end{proof}

It follows that quantifying a sole statement requires a symmetric nugget $U_k$, which implies by lemma \ref{nuggetslemma} the matrices in $U_k$ are $M_t$ and $M_t^{'}$, as well as that $U_l=U_m^{'}$, which implies $U_l = M_vM_w^{'}$ and $U_m=M_wM_v^{'}$. In those cases, one and only one configuration of nuggets (and matrices) exists, since transposing equals the trivial identity permutation. If $U_k=U_q^{'}$ for any $U_k=M_t \times M_s^{'}$ and $U_q=M_s \times M_t^{'}$ will contain the same set of values, but in a different order. Similarly, in any $U_lU_m$ that has $U_sU_r$ as a distinguishable transpose, all values $U_lU_m$ uniquely exist as transpose in $U_sU_r$, and vice verse. So for any $U_l = M_vM_w^{'}$ and $U_m=M_xM_y^{'}$ there exists $U_r = M_wM_v^{'}$ and $U_s=M_yM_x^{'}$, such that $\sum_{k}(M_{v_{ik}}M_{w_{ki}})(M_{x_{kj}}M_{y_{jk}}) = \sum_{k}(M_{y_{jk}}M_{x_{kj}})(M_{v_{ik}}M_{w_{ki}})\,\forall\, \{i,j,k\}$. Hence, only when either or both variables in $\rho$ have a distinguishable transpose a distinguishable twin statement exists that produces the exact same value for $\rho$, i.e. showing duality for this subset of balance correlations.

\subsection*{Cardinality}
Definition \ref{BC} determines any balance correlation $\rho(U_k, U_lU_m)$ is a bi-variate Pearson correlation.    This leads us to consider two conditions, or distinguishing assumptions, as special cases.  Although we recognise more complex situations could apply, these two limiting conditions will provide an upper and lower cardinality bound for $\mathfrak{B}$.

\begin{assumption}\label{sym_assumption}
All matrices in $\mathfrak{M}$ are symmetric.
\end{assumption}

This implies that for all nuggets in $\mathfrak{Z} = \{M_r \times M_{s}^{'} \, | \, r \neq s, M_r \in \mathfrak{M}, M_{s}^{'} \in \mathfrak{M}^{'}\}$ the variances $\var(Z_h)=0$.  Since the Pearson correlation is defined if and only if $\var(U_k)>0$ and $\var(U_lU_m)>0$,  the balance correlations that include $Z_h$ are not defined, and cannot be considered. Therefore, under assumption \ref{sym_assumption} the nugget set cardinality will be $|\mathfrak{U}|=m$, and will only contain symmetric nuggets.

The symmetry of the nuggets implies there are $m$ possible nuggets for the first variable in $\rho$. Because, there is no restriction on the use of nuggets in any of the $2$ positions in the second variable, the number of possible combinations to constitute it, in general, is given by 

\begin{equation}\label{BCcardinalitysymm}
\eta=\Biggl(\binom{m}{2}\Biggr)=\frac{(m-1+2)!}{(m-1)!2!},    
\end{equation}
where $m$ is the number of different nuggets in $\mathfrak{Z}$. It follows that $|\mathfrak{B}|= \eta \cdot m$ is the cardinality of the largest possible set of unique balance correlations. 

In the specific case discussed in  this paper, where we are limiting ourselves to a set of three symmetric relations (\textbf{P}, \textbf{N}, and \textbf{Z}), $\eta=6$ and thus the largest possible set of \textit{unique} correlations $=3\cdot 6 = 18$.  Therefore, of the $m^3 =27$ possible balance correlations, only 18 of them will be unique, and the remaining nine will be twins, paired with nine elements in this set of 18 unique balance correlations.  With nine pairs of twins, that means the remaining nine balance correlations will have no twins.

\begin{assumption}\label{asym_assumption}
    All matrices in $\mathfrak{M}$ are asymmetric, nuggets have $\var(U_k)>0\, \forall \,k \in \{1,...,g\}$, and the inner product of nuggets have $\var(U_lU_m)>0\, \forall \,l, m \in \{1,...,g\}$.
\end{assumption}

The proof of theorem \ref{theorem.1} provides the conditions under which a specific configuration of nuggets doesn't have a twin. It requires a symmetric nugget in the first variable in $\rho$, and a symmetric inner product of nuggets in the second variable. The latter condition can only hold when no more than two different matrices constitute the two nuggets. The order of the two matrices do define the specific nuggets, and hence is important. However, these two nuggets also need to be transposes of each other. Given $m$ asymmetric matrices there are exactly $m$ symmetric nuggets, according to the proof of lemma \ref{nuggetslemma}. Furthermore, there are exactly $g=m^2$ nuggets, which includes their transposes. So, the total number of possible inner products equals $m^4$, of which $m^2$ will be inner products with the transpose. Hence, there are $\alpha = m \cdot m^2 = m^3$ number of configurations of matrices in the variables of $\rho$ that quantify a sole statement. The other possible balance correlations all have exactly one other equal quantification of a twin statement. Therefore,

\begin{equation}\label{BCcardinalityasymm}
|\mathfrak{B}|=\frac{1}{2}(m^6- m^3)+ m^3=\frac{1}{2}m^3(m^3+1)
\end{equation}
where $m^6$ is the number of combinations of all possible nuggets (2 matrices in a nugget) in the three nuggets used in $\rho(U_k, U_lU_m)$. For any set of mixed symmetric and asymmetric matrices the cardinality of $|\mathfrak{B}|$ will be between those in Eq~\ref{BCcardinalitysymm} and Eq~\ref{BCcardinalityasymm}.

The logical entanglement of statements is both interesting and limiting this approach. In fact it can be solved by extending the approach to consider an extra temporal dimension to the data. Note however, that the balanced states all rely on the assumption of reciprocal relations (symmetric nuggets). As such, at any moment in time these states, according to the theory, may be expected to occur more than unbalanced states. This motivates the use of balance correlation as a cross-sectional concept, and to rely on a subset of $|\mathfrak{B}|$, which reflects the statement. Yet, from an exploratory perspective it remains of interest to inspect the values of other measures as this can reveal information about the process of balance formation.  

\section*{Acknowledgments}
We like to thank participants at Networks 2021 (Bloomington, USA), Sunbelt 2022, (Cairns, Australia), Sunbelt 2023 (Portland, USA), ION IX (Lexington, USA), and EUSN 2023 (Ljubljana, Slovenia) for valuable remarks.

%\bibliographystyle{abbrvnat}
%\bibliography{references, references2, structuralleverage, TCbib}

\end{document}